\documentclass[10pt,final,journal]{IEEEtran}
\pdfoutput=1

\usepackage[ruled,vlined]{algorithm2e}
\usepackage{color}
\usepackage{epsfig}
\usepackage{url}
\usepackage{cite}
\usepackage[center]{subfigure}
\usepackage{amsmath,lipsum}
\usepackage{bm}
\usepackage{epstopdf}
\usepackage{amsfonts}
\usepackage{flushend}

\providecommand{\eat}[1]{}
\begin{document}
\title{\huge\bfseries {Practical Interpolation for Spectrum Cartography through Local Path Loss Modeling}}
\author{Shweta Sagari, Larry Greenstein, Wade Trappe\\
\{shsagari, ljg, trappe\}@winlab.rutgers.edu\\
WINLAB, Rutgers University, North Brunswick, NJ 08902, USA.}
% An Interpolation Scheme for Constructing Radio Frequency Maps from Spatial Samples
% Two Tiered Approach for Constructing Radio Coverage Maps from Theory to Practice

\maketitle

\begin{abstract}
A fundamental building block for supporting better utilization of radio spectrum involves predicting the impact that an emitter will have at different geographic locations.
To this end, fixed sensors can be deployed to spatially sample the RF environment over an area of interest, with interpolation methods used to infer received power at locations between sensors. This paper describes a radio map interpolation method that exploits the known properties of most path loss models, with the aim of minimizing the RMS errors in predicted dB-power. We show that the results come very close to those for ideal Simple Kriging. Moreover, the method is simpler in terms of real-time computation by the network and it requires no knowledge of the spatial correlation of shadow fading. Our analysis of the method is general, but we exemplify it for a specific network geometry, comprising a grid-like pattern of sensors. We also provide comparisons to other widely used interpolation methods.
\eat{
Unfortunately, it is rarely possible to densely deploy sensors and thus obtaining a full map of radio frequency (RF) power levels requires tools that use a collection of spectrum measurements to infer the expected power levels at locations where there is no measurement infrastructure. Building a radio map at a granularity more refined than the deployed sensors requires signal interpolation tailored to the characteristics of RF propagation. This paper explores an interpolation scheme based that incorporates path loss modeling to accurately estimate the power levels at locations bounded by four spectrum sensors deployed in a rectangular pattern.
}
\end{abstract}


\section{Introduction}
\label{sec:intro}

There are several hurdles facing the vision of open and dynamic sharing of spectrum\cite{Mandayam:08}. Perhaps the foremost challenge is the need to create a database of geographical radio maps with information such as location, frequency, and perceived radio power levels\cite{Google_database,Microsoft_observatory}. To build such a detailed radio map, radio power from emitters can be measured using a distributed collection of spectrum sensors that span a region of interest. The set of sensor measurements can be processed at a logically central database and interpolated over the large area to build the radio map. The use of radio maps can be extended to applications such as spectrum usage policing, network planning and deployment, and informing spectrum marketplaces.

In this paper, we examine the problem of constructing a radio map that contains estimates of the RF received power in a specified frequency band over a network's coverage area. As a starting point for our analysis, we assume that there is a single emitter, and all sensors scan the same frequency band, where each of the sensors measures and reports the power received from the emitter. We initially assume, moreover, that the emitter location is known. The position of an emitter can be determined, for example, using any of the numerous emitter localization methods that are common practice \cite{Ziskind1988_Maximum,Misra2011_Global,Langendoen2003_Distributed}. By studying this case in depth, we will uncover results that permit the cases of unknown emitter location and multiple emitters to be treated as well.

We begin the paper in Section \ref{sec:related} by briefly summarizing the relevant literature. In Section \ref{sec:approach}, we describe the propagation model associated with typical outdoor wireless environments, and show how it suggests a possible approach to optimal interpolation, which we call the Stochastic Method (\textit{SM}). In Section \ref{sec:SM0}, we derive the root-mean square (RMS) interpolation error for SM under ideal circumstances (we call this case \textit{SM-0}) and then show it to be equivalent to a basic form of Kriging, which is referred to as Simple Kriging\cite{Cressie1990_origins}. Since SM-0 requires knowledge that might not be available in practice, we then discuss two approaches for reducing SM to practice: SM-1, which is possible but requires special knowledge of the propagation environment and considerable real-time calculation; and SM-2 which requires neither.  In Section V, we quantify the RMS errors for a particular sensor network geometry, showing that SM-1 and SM-2 are very close to each other in performance and only slightly less accurate than the ideal SM-0. We also compare these algorithms with other traditional interpolation methods, and show that SM-2 gives the best tradeoff between accuracy and simplicity. We summarize our findings and their extensions in Section VI, and conclude the paper in Section VII.
\section{Related Work}
\label{sec:related}

Spectrum mapping or spectrum cartography has been applied to different aspects of wireless networks, ranging from estimating radio coverage to wireless network planning\cite{Robinson2008_assessment}, mesh/multi-hop networks\cite{Agrawal2009_correlated}, radio tomographic imaging\cite{Wilson2010_RTI}, cognitive radio networks \cite{Zhao2007_applyREM,Feki2008_CRNcartography,MateosGiannakis2009_spline}, and LTE carrier aggregation\cite{Dwarakanath2013_modeling}. A related application of spectrum mapping involves the  need to detect anomalous emitters and to quantify their impact on radio power levels across a geographic area. As such, there are many different areas of related work, including spectrum scanning, signal detection in complex radio environments, and multi-dimensional interpolation.  In the area of spectrum scanning, recent work such as \cite{GarnaevScan}, explore scheduling a spectrum sensor's time-frequency scanning across a large swath of bandwidth in hopes of detecting an anomalous signal. The problem of detecting signals in a variety of fading environments (including Rayleigh, Rican and Nakagami channels) was explored in \cite{Digham:03}. To combat the uncertainties caused by fading channels in performing signal detection and to support the estimation of a emitter's impact spatially, it is desirable to use multiple sensors cooperatively.  For example, in \cite{liu2009aldo}, the authors examine an approach that leverages measurements across an area to decide whether the signal pattern across the region suggests the presence of an anomalous signal, while \cite{Visotsky:05} explores collaborative detection of primary user TV signals in dynamic spectrum access (DSA).

Over the past few decades, radio path loss prediction has been studied using a variety of methods, such as theoretical models, stochastic fading models, interpolation/mapping methods, measurement based/correlated models, and geostatistical models \cite{Phillips2013_SurveyWirelessPL}. Basic path loss and stochastic fading models assume prior knowledge and can perform well in specific circumstances. However, they are costly in terms of the time and data acquisition needed to tune them accurately to address other cases \cite{Phillips2011_bounding}. On the other hand, measurement-based models like interpolation and geostatistical methods use measurements from active sensor nodes to estimate physical parameters for given circumstances. In our study, we introduce a practical version of this approach to radio mapping that involves local estimation of path loss parameters to ultimately arrive at better radio map interpolation.

In previous work, global interpolation techniques - Kriging and thin-plate splines - have been studied for spectrum cartography. Under these techniques, measurements from the entire set of sensors are collected and processed at a central location. Kriging is a reliable approach based on the best linear unbiased estimator (BLUE) \cite{Cressie1990_origins}. Its variants, such as Kriged Kalman filter (universal Kriging) \cite{AneseGiannakis2011_chGain}, and ordinary Kriging \cite{Ureten2012_comparisonCartography,Feki2008_CRNcartography} have been widely studied. But these assume first- and second-order stationarity of the spatial data, which is not generally applicable in realistic environments. The use of thin-plate splines, on the other hand, is based on radial basis functions and does not make any channel assumptions. While this approach performs well, even in the absence of precise frequency and bandwidth information, unfortunately the computational complexity increases enormously with the number of sensors and basis functions \cite{Ureten2012_comparisonCartography,BazerqueGiannakis2011_GroupLasso,Bazerque2011_basisPur,MateosGiannakis2009_spline}.

In contrast, local spatial interpolation techniques consider only those sensors and their measurements that neighbor the location of interest, and therefore deal well with local changes or variations in the environment, as well as with directional antennas used at emitters. Amongst the local interpolation techniques, Nearest Neighbor (NN), linear interpolation, and Natural Neighbor (NaN) are simple to implement. Unfortunately, while simple to implement, these methods may not be suitable when the characteristics underlying the data changes quickly between nodes or because sensors are often only sparsely positioned\cite{Bolea2011_RecSignal,Ledoux2005_NaturalNeighbour}. Inverse Distance Weighting (IDW), which is a close relative to the approach we present in this paper, offers a simple and effective form of local  spatial interpolation \cite{Angjelicinoski2011_Comparative,Dwarakanath2013_modeling}, but does not involve local path loss parameter estimation.

\section{Radio Mapping Approach}
\label{sec:approach}

The mapping approach described here exploits the mathematical structure of most terrestrial path loss models, which are based on numerous  measurement and modeling campaigns over many years. To be specific, the majority of path loss models published for outdoor environments are of the form $PL(d) = B + \Gamma log(d/d_r) + S$, where $B$ and $\Gamma$ are constants that depend on frequency, antenna heights and gains, and terrain details; $S$ is the statistical variation of path loss about $[B + \Gamma log (d/d_r)]$ over all Tx-Rx separations of distance $d$. Here, $d_r$ is a reference distance, which we will assume, for convenience only, to be 1 m. Path loss models having this form can be found, for example, in \cite{Phillips2011_bounding, Hata_model,  Erceg1999_empiricalPL, Phillips2013_SurveyWirelessPL, Rappaport1996_wireless,Goldsmith2005_wireless}. It is customary to regard $[B + \Gamma log(d/d_r)]$ as the median path loss for distance $d$, and $S$ as shadow fading, typically modeled as a zero-mean Gaussian random process over the environment. Finally, the received power at a given point on the terrain can be written as, $P_r = P_t + PL(d)$, which is the quantity to be mapped. Here, $P_r$ corresponds to receive power, and $P_t$ is the transmit power. Invoking the generic path loss form assumed here, $P_r$ is
\begin{equation}
\label{eq:PL}
P_r = A + \Gamma \log(d/d_r) + S,
\end{equation}
where $A = P_t + B$.

The constants $A$ and $\Gamma$ are context-specific in that they depend on the transmitted power and the terrain features over the areas to be mapped. In an environment filled with sensors, they can be computed by the network for any small (local) area by measuring received power at $n$ nearby sensors and performing least-squares estimation (LSE) or other forms of estimation\cite{Bickel2001_mathematical,Rice1995_mathematical}\footnote{Throughout this paper we shall explore the use of least-squares estimation, due to its combination of simplicity and good performance, but note that our methodology can apply equally well to other approaches, such as maximum-likelihood estimation.}. With $A$ and $\Gamma$ thus quantified, the median received power
\begin{equation}
\label{eq:PLm}
P_{m} = A + \Gamma \log(d/d_r).
\end{equation}
can then be computed for any given point within the local area, and thus only $S$ needs to be further estimated at that location of interest. This can be achieved by measuring $S$ at each of the $n$ nearby sensors (i.e., by subtracting the median at the sensor from the measured power) and then forming an $n$-fold weighted sum of the resulting $S$-estimates. This is the essence of our approach, which we describe in detail in later sections, and can be applied to any and all points within the coverage area to create the radio map.

For finite $n$, the estimates of $A$ and $\Gamma$ will be imperfect due to the corrupting effects of the random shadow fading, with the estimates tending to improve as $n$ increases. We require that $n > 2$ in all cases and later in this paper will examine a specific scenario wherein $n = 4$. We will also propose a simple weighting scheme for estimating $S$ at a given point (which yields a powerful variant of inverse distance weighting, IDW) that does not require knowing the spatial statistics of shadow fading; and we will see that, in terms of RF power estimation accuracy, the results are close to a best-case bound, which we will derive.

\eat{
\subsection{Conventional Radio Path Loss Model}
\label{subsec:PLModel}
We will use the convention that path loss is the dB ratio of received power to transmitted power. Thus, the dB received power at a given point on the terrain from an emitter $d$ meters away is given by
\begin{equation}
\label{eq:RecPwr}
P_r = P_t + PL(d),
\end{equation}
where $P_r$ and $P_t$ are the received and transmitted powers, respectively, in consistent units (e.g., dBm); and $PL(d)$ is the path loss, including the effects of gains in the transmit and receive antennas.

The mapping approach described here exploits the mathematical structure of most terrestrial path loss models, based on countless measurement/modeling campaigns over many years. To be specific, the majority of path loss models published for outdoor environments are of the form
\begin{equation}
\label{eq:PL}
PL(d) = C + 10\gamma \log(d/d_r) + s,
\end{equation}
where $C$ and $\gamma$ are constants that depend on frequency, antenna heights and gains, and terrain details; $d_r$ is a reference distance (chosen as $d_r = 1$m); and $s$ is the statistical variation of path loss about $[C + 10\gamma\log (d/d_r)]$ over all Tx-Rx separations of distance $d$. Path loss models having this form can be found, for example, in \cite{Phillips2011_bounding, Erceg1999_empiricalPL}. It is customary to regard $[C + 10\gamma \log(d/d_r)]$ as the median path loss for distance $d$, and $s$ as the shadow fading, typically modeled as a zero-mean Gaussian random variable over the environment. Finally, the received power, $P_r = P_t + PL(d)$, which is the quantity to be mapped, can be modeled as $Pr = A + 10\gamma \log(d/dr) + s$, where $A = Pt + C$. Thus, we modify the median path loss equation as
\begin{equation}
\label{eq:PLm}
PL_m = A + 10\gamma \log(d/d_r).
\end{equation}

Shadow fading $s$ is a log-normal Gauss-distributed with zero mean and standard deviation $\sigma$ [dB]. Its variability over the terrain can be described by a 2-dimensional random process proposed in Gudmundson's model which is measurement supported\cite{Gudmundson1991_Correlation} and widely accepted\cite{Zhao2004_Path,Agrawal2009_correlated}. Spatial correlation among $s$-values at locations $a$ and $b$ is defined by
\begin{equation}
\label{eq:auto-co}
c_{ab} = \sigma^2 \exp\left(-\frac{d_{ab}}{X_c}\right).
\end{equation}
where $X_c$ is the shadow fading correlation distance and $d_{ab}$ is the distance between $a$ and $b$.  The values of both $\sigma$ and $X_c$ are environment-specific\cite{Gudmundson1991_Correlation,Goldsmith1994_Error}.

\subsection{Local Path Loss Model}
The function constants $A$ and $\gamma$ in (\ref{eq:PL} and \ref{eq:PLm}) are context-specific, in that they depend on the transmitted power and the terrain features over the area to be mapped. Hence, they must be determined for each specific area. In an environment filled with sensors, this is made possible by measuring received power at n nearby sensors and doing least-squares estimation. With $A$ and $\gamma$ thus quantified, the component $[A + 10\gamma log(d/do)]$ can be computed for any given point within the area, and so only $s$ needs to be further estimated at that point. This can be done by measuring $s$ at each of the $n$ sensor locations and then forming an $n$-fold weighted sum of these $s$-values.

Furthermore, in a practical setting, shadow-fading is highly correlated over small to moderate distances and decorrelates over large distances\cite{Agrawal2009_correlated}. Thus, the path loss at a particular point is best estimated using the known (or measured) path loss values from the set of sensors located nearest to that point. Thus, we specify a \emph{Local Path Loss} model which gives a more accurate estimate of the path loss for a given emitter for a given smaller area and its surrounding.

As an example, we consider an arbitrary point $P_0$ located in a area bounded by $n$ sensors. At $P_0$ shadow fading $s_0$ includes the bias factor $Z_n$ which appears due to correlation of $s_0$ with shadow fading $[s_1\;s_2\;..\;s_n]^T$ at sensors. Thus, we propose to include bias $Z_n$ with the constant parameter $A$ in Eq.~\ref{eq:PLm} and problem gets deduced to the estimation of $A+ Z_n, \gamma$, and $s"_0 = s_0 - Z_n$ at $P_0$. The proposed local path loss model is
\begin{equation}
\label{eq:lo_PL}
\begin{aligned}
PL_0 = A^" + 10\gamma\log d_0 + s^"_0,\\
\mbox{where } A^" = A + Z_n \mbox{ and }
s^"_0 = s_0 - Z_n.
\end{aligned}	
\end{equation}
In practice, finding $Z_n$ is difficult without any a priori joint distribution information of $[s_1\;s_2\;..\;s_n]^T$. Thus we approximate $Z_n$ as
\begin{equation}
\label{eq:Z_n}
Z_n = \frac{1}{n} \sum_{i=1}^{n} s_i.
\end{equation}
Note that values of $s^"$ at sensors average to $0$, as desired in a model constructed from these sensor measurements. For the local path loss model, the median path loss at point $P_0$ is redefined as
\begin{equation}
\label{eq:rev_medPL}
PL^"_{m_0} = A^" + 10\gamma\log d_0.
\end{equation}
}

\section{The Stochastic Method}
\label{sec:SM0}

In this section, we present the heart of our approach to performing the underlying interpolations associated with estimating received power. We start by first providing a quick background discussion regarding the spatial characteristics of shadow fading, then move to presenting the idealized form of our interpolation approach, which includes analyzing the first and second moments of shadow fading at an arbitrary point.
%and then determine the underlying first and second order statistics for shadow fading at an arbitrary point.

\subsection{Spatial Correlation of Shadow Fading}

The interpolation approach that we will describe will involve: (i) using in-field measurements  to estimate the `deterministic' part of $P_r$ at a given point; and (ii) focus on estimating the `random' part, $S$, at the point of interest. One can envision the shadow fading component as a two-dimensional stochastic process over the terrain, where $S$ at any point is a Gaussian random variable of zero mean and standard deviation $\sigma$; and the relationship between $S$-values at any two points $i$ and $j$ can be characterized by an autocorrelation function,
\begin{equation}
\label{eq:auto-co_1}
<S_i S_j> = c_{ij},
\end{equation}
where $<X>$ denotes the expected value of $X$\eat{, and $c_{ij}$ is a correlation}. The optimal way to estimate $S$ at a location between measurement points (sensors) is to know and exploit the correlation properties of $S$ over the terrain, and hence we call this the stochastic method (SM).

A popular formulation for $c_{ij}$ that is simple to use and supported by data in the literature \cite{Gudmundson1991_Correlation}, is the decaying exponential,
\begin{equation}
\label{eq:auto-co}
c_{ij} = \sigma^2 \exp\left(-\frac{d_{ij}}{X_c}\right),
\end{equation}
where $d_{ij}$ is the physical distance between points $i$ and $j$, and $X_c$ is the so-called correlation distance of shadow fading on the terrain. The value of this parameter depends on the type of terrain, and empirical results have been reported for different environments \cite{Goldsmith1994_Error,Zhao2004_Path,Agrawal2009_correlated}.

\eat{We define, for later use, the $(n+1) \times (n+1)$  correlation matrix $C$ whose $(i,j)$-th element is $c_{ij}$, with $i$ and $j$ ranging from $0$ to $n$.}
In our computations of RMS interpolation error, we will invoke the above correlation coefficient as well as others, showing that the precise shape of the function is not a first-order concern in the underlying problem.

\subsection{The Ideal Case: SM-0}
Assume that $P_r$ is to be estimated at a given point on the terrain (labeled as point 0), which is surrounded by $n$ measuring sensors $(n > 2)$. The parameters $A$ and $\Gamma$ are estimated from the $n$ measurements of received power, and will be imperfect estimates due to the $S$-values, which act like additive noise. To obtain a theoretical best-case accuracy, however, we assume at first that these estimates are perfect. Therefore, the median value of $P_r$ at point 0 is exact, and the network need only predict $S_0$ ($S$ at point 0). To quantify the minimal RMS error in this prediction, we use the mathematics of multivariate Gaussian distributions: Assuming that $S_1,..,S_n$ are measured precisely at the sensors, $S_0$ is modeled as
\begin{equation}
\label{eq:mu0_sig0}
[S_0|S_1,..,S_n] \simeq \mu_0 + \sigma_0 u,
\end{equation}
where $\mu_0$ is the mean of $S_0$ conditioned on $S_1,..,S_n$ and is a weighted sum over these $S$-values; $u$ is a zero-mean, unit-variance Gaussian random variate; and $\sigma_0$ is the standard deviation of the variation about the mean. We will show that the weights over the $n$ $S$-values can be determined if its correlation is known. Thus, in the ideal case, the expected value of $S_0$ can be known, in addition to the median of $P_r$. This leaves only $\sigma_0 u$ as the unknowable component of $P_r$. Thus, $\sigma_0$ is the irreducible RMS error in interpolating $P_r$ from the sensor measurements.
\eat{
From multivariate Gaussian statistics, this can be shown to be
\[\sigma_0^2 = \sigma^2 - \bm{c_0^T C_n^{-1} c_0}.\]
}

It is worth noting that the above approach is equivalent to a basic form of Kriging, which is often referred to as Simple Kriging\cite{Cressie92_statSpati}. We show in Section \ref{app:equiSMKrig} of the Appendix that the minimum RMS error is equivalent to the RMS error obtained from Simple Kriging, where the bias term is precisely known.

\subsection{Determining $\mu_0$ and $\sigma_0$}
\label{subsec:est_weights}
For the ideal case, SM-0, the environmental parameters $A$, $\Gamma$, and shadow fading values at $n$ sensors are perfectly known. Thus, $P_{r,0}$ at point 0 is given as
\begin{equation}
\label{eq:Pr0}
P_{r,0} = A + \Gamma \log (d_0/d_r) + [\mu_0 + \sigma_0 u],
\end{equation}
where $d_0$ is the distance from point 0 to the emitter, and the bracketed term corresponds to the shadow fading component, (\ref{eq:mu0_sig0}).

%here $\bm{W}^T\bm{S}$ is a weighted sum of shadow fadings at $n$ sensors with vectors $\bm{W} = [w_1\;w_2\;..\;w_n]^T$ for weights and $\bm{S}=[S_1\; S_2\; \dots\; S_n]^T$.
%Due to the random properties of shadow fading, there is a Gaussian spread of possible values for a given $\bm{S}$-vector at each given point 0. Therefore, no solution is available to calculate the ideal assignment of weights, even if shadow fading values at sensors are known perfectly. We, thus, use the multivariate Gaussian distribution of $S$-values to minimize the RMS error for estimation of shadow fading $S_0$ at the point 0.

Under the ideal conditions assumed, we can determine $\mu_0$ and $\sigma_0$ exactly, since $\mathbf{S} = [S_1, \cdots, S_n]$ is an $n$-fold set of zero-mean Gaussian variates of known correlation matrix. The joint probability density function (pdf) of this set is \cite{Yates1999_stochastic}
\begin{equation}
\label{eq:shadow_PDF}
\begin{aligned}
f_S(s) = \left((2\pi)^n|\bm{C_n}|\right)^{-1/2} \exp\left(-\frac{1}{2}(\bm{S^T C_n S})\right)
\end{aligned}
\end{equation}
where $\bm{C_n}$ is $n\times n$ correlation matrix of $\bm{S}$ with determinant $|\bm{C_n}|$ and each of its elements is computed by (\ref{eq:auto-co_1}). Thus, from multivariate Gaussian statistics\cite{Eaton1983_multiVar}, $[S_0 | \mathbf{S}]$ has mean $\mu_0$ and standard deviation $\sigma_0$ given by
\begin{equation}
\label{eq:mu5}
\begin{aligned}
\mu_0 = (\bm{c_0^T C_n^{-1}})\bm{S},\\
\sigma_0^2 = \sigma^2 - \bm{c_0^T C_n^{-1} c_0},
\end{aligned}
\end{equation}
respectively, where $\bm{c_0}$ is the $n \times 1$ cross-correlation vector of $S_0$ with $\bm{S}$, where the $j$-th element is given as $c_0(j) = <S_0 S_j>$.
%We note that $\sigma_0$ is the minimized RMS estimation error and as the number of sensors $n \rightarrow \infty$, $\sigma_0^2=0$.
From (\ref{eq:mu5}), we see that $\mu_0$ is a weighted sum over the $S$-values at the $n$ sensors, $\mu_0  = \bm{W^TS}$, where the weight vector is
%The optimal value for the weights associated with estimating $\mu_0$ from a given collection of $\bm{S}$ are given by $\bm{W}$ as
\begin{equation}
\label{eq:W}
\bm{W} = \bm{(c_0^T C_n^{-1})^T}.
\end{equation}

%Matrix $\bm{c_0 C_n^{-1}}$ is the matrix of regression coefficients and matrix $(\sigma^2 - \bm{c_0^T C_n^{-1} c_0})$ is the Schur complement of $\bm{C_n}$ in the covarince matrix of
%$C = \begin{bmatrix}
%       \sigma^2 & \bm{c_0}\\[0.3em]
%       \bm{c_0^T} & \bm{C_n}
%     \end{bmatrix}$.

Table~\ref{tab:notation} lists notations which we have used here and in the rest of the paper.
\begin{table}[t]
\caption{Notation description}
\centering
\begin{tabular}{| l | l |}
\hline
\textbf{Notation} & \textbf{Parameter}\\
\hline
$n$ & Number of sensors\\
\hline
$A^{"}$ & Redefined path loss constant\\
\hline
$\gamma$ & Path loss exponent\\
\hline
$P_{r,i}$ & True measurement (received power) at $i$\\
\hline
$P_{m,i}$ & True median power at $i$\\
\hline
$\sigma$ & Standard deviation of shadow fading\\
\hline
$X_c$ & Shadow fading correlation distance\\
\hline
$S_i$ & True shadow fading at $i$\\
\hline
$\bm{W}$ & Weights vector assigned to sensor measurements\\
\hline
$d_i$ & distance between emitter and $i$\\
\hline
SM & Stochastic Method\\
\hline
\textit{RMSE} & Root Mean Square Error\\
\hline
\end{tabular}
\label{tab:notation}
\vspace{-0.2cm}
\end{table}

\eat{
\subsection{Estimation of Weights}
Firstly, we employ a methodology to compute a lower bound on the path loss estimation error which can be achieved theoretically when path loss measurements from $n$ sensors are available. For this purpose, we follow the following procedure:
\begin{enumerate}
\item We assume an ideal environment where values of $A$ and $\gamma$ are perfectly known.
\item From equations \ref{eq:PL} and \ref{eq:PLm}, shadow fading values at $n$ sensors $\bm{S} = [S_1\; S_2\; \dots\; S_n]$ are also known perfectly.
\item Therefore, path loss at arbitrary point $P_0$ is evaluated as
\begin{equation}
\label{eq:PL0_SM0}
\begin{aligned}
PL_0 = A + 10 \gamma \log d_0 + \bm{W}^T\bm{S},
\end{aligned}
\end{equation}
where $\bm{W}^T\bm{S}$ is a weighted sum of shadow fadings at $n$ sensors with weights vector $\bm{W} = [w_1\;w_2\;..\;w_n]^T$.
\end{enumerate}

No solution is available to calculate the ideal assignment of weights, even if shadow fading values at sensors are known perfectly. Due to the random properties of shadow fading, for a given $\bm{S}$-vector, there is a Gaussian spread of possible values at each virtual point, with a mean and standard deviation we can quantify. Thus, we exploit  the multivariate Gaussian distribution of $S$-values, as given in following sub-section, to minimize RMS error for the estimation of shadow fading $S_0$ at $P_0$.

Shadow fading values at sensors, $\bm{S}$, are zero-mean Gaussian distributed, thus, they are jointly Gaussian with their probability distribution function (PDF) as\cite{Yates1999_stochastic}
\begin{equation}
\label{eq:shadow_PDF}
\begin{aligned}
f_S(s) = [(2\pi)^n|\bm{C_n}|]^{-1/2} \exp(-\frac{1}{2}(\bm{S^T C_n S}))
\end{aligned}
\end{equation}
where $\bm{C_n} \in \mathbb{R}^{n\times n}$ is covariance of $\bm{S_n}$; $|\bm{C_n}|$ is the determinant of $\bm{C_n}$ and each element of $\bm{C_n}$ is computed using Eq.~\ref{eq:auto-co}.

The value of $S_0$ at $P_0$ conditioned on the known $\bm{S}$ at sensors is given as
\[[S_0|\bm{S}] \simeq \mu_0 + \sigma_0u,\]
where $[s_0|\bm{s}]$ is a random quantity with a mean $\mu_0$ and a Gaussian variation about $\mu_0$ with standard deviation $\sigma_0$. $u$ is a zero-mean, unit-variance Gaussian. From multivariate zero-mean Gaussian distribution statistics\cite{Eaton1983_multiVar}, we have mean and variance as
\begin{equation}
\label{eq:mu5}
\begin{aligned}
\mu_0 = (\bm{c_0^T C_n^{-1}})\bm{S},\\
\Sigma_0^2 = \sigma^2 - \bm{c_0^T C_n^{-1} c_0},
\end{aligned}
\end{equation}
respectively, where $\bm{c_0} \in \mathbb{R}^n$ is a cross-covariance vector of $S_0$ and $\bm{S}$; and $\sigma^2$ is the variance of $s_0$. Matrix $\bm{c_0 C_n^{-1}}$ is the matrix of regression coefficients and matrix $(\sigma^2 - \bm{c_0^T C_n^{-1} c_0})$ is the Schur complement of $\bm{C_n}$ in the covarince matrix of
$\begin{bmatrix}
       \sigma^2 & \bm{c_0}\\[0.3em]
       \bm{c_0^T} & \bm{C_n}
     \end{bmatrix}$
 $\in \mathbb{R}^{n+1 \times n+1}$ representing covariance of $s_0$ with $s_1$ through $s_n$. We also note here that as number of sensors $n \rightarrow \infty$, $\sigma_0^2=0$.

Here, $\mu_0$ is the least-squares estimate of $s_0$ and $\sigma_0$ is the minimized RMS estimation error. Thus, the optimal for weighting for elements of $\bm{S}$ are given by $\bm{W}$ as
\begin{equation}
\label{eq:W}
\bm{W} = \bm{(c_0^T C_n^{-1})^T}.
\end{equation}

In summary, path loss estimation at $P_0$ and its RMS error is given as
\begin{equation}
\label{eq:PL0_SM0_fin}
\begin{aligned}
PL_0 = A + 10 \gamma \log d_0 + (\bm{c_0^T C_n^{-1}})\bm{S}\\
\Sigma_0 = \sqrt{\sigma^2 - \bm{c_0^T C_n^{-1} c_0}},
\end{aligned}
\end{equation}
respectively.

\subsection{Equivalence of SM-0 and Kriging}
Kriging is a spatial optimal linear predictor which has a form of weighted averaging. Weights for Kriging are chosen based on best linear unbiased estimator (BLUE) where weights depend upon locations of sensors used in the prediction process and covariannce among them. Kriging is widely adopted in literature [references] for different interpolation applications with three different variations - (1) simple Kriging, (2) ordinary Kriging, and (3) Universal Kriging [references].

In our application for ideal case with known $A$ and $\gamma$, we will consider simple Kriging for which mean of random process is assumed to be known. Optimal path loss estimation at $P_0$, $Z^*(P_0)$, is given by
\begin{equation}
\label{eq:simKrigPL}
Z^*(P_0) = \bm{c_0^T C_n^{-1}PL} + (1 - \bm{c_0^T C_n^{-1}1})\mu,
\end{equation}
where $PL \in \mathbb{R}^n$ is a path loss measurements at $n$ sensors, $\bm{1} \in \mathbb{R}^n$ is a vector of \textit{1}s, $\mu$ is a known mean of random path loss process. Here, $\mu$ is equivalent to median path loss term defined by Eq. \ref{eq:rev_medPL} as $PL_{m,i} = A + 10\gamma\log d_i$ at sensor $i$ where $d_i$ is the distance between $i$ and emitter and $A$ and $\gamma$ are known. It is clear that median path loss at each sensor is based on $d_i$ and can not be assumed to be a single $\mu$ values as given by Eq.(\ref{eq:simKrigPL}). Thus, we propose following modifications for Simple Kriging for path loss prediction as follows:
\begin{equation}
\label{eq:simKrigPL_rev}
Z^*(P_0) = \bm{c_0^T C_n^{-1}PL} + (PL_{m,0} - \bm{c_0^T C_n^{-1}PL_m}),
\end{equation}
where $PL_{m,0}$ is a median path loss at $P_0$ and $\bm{PL_m} \in \mathbb{R}^n$ is vector of median path loss at sensors.

If we expand Eq. \ref{eq:simKrigPL_rev} using equations \ref{eq:PL} and \ref{eq:PLm}, we see that \ref{eq:simKrigPL} and \ref{eq:simKrigPL_rev} are exactly same. Thus, we conclude that SM-0 and simple Kriging are equivalent in the ideal case and gives the lower bound on path loss estimation error. Thus, in the rest of the paper, we will take SM-0 as a baseline to study non-ideal (practical) scenarios where $A$ and $\gamma$ are unknown. We list the notations used throughout the paper in Table~\ref{tab:notation}.

\begin{table}[t]
\caption{Notation description}
\centering
\begin{tabular}{| l | l |}
\hline
\textbf{Notation} & \textbf{Parameter}\\
\hline
$n$ & Number of sensors\\
\hline
$A$ & Path loss constant\\
\hline
$\gamma$ & Path loss exponent\\
\hline
$\bm{PL} \in\mathbb{R}^n, PL_i$ at $i$ & Actual path loss measurement\\
\hline
$\bm{PL_m} \in\mathbb{R}^n, PL_{i,m}$ at $i$ & Actual median path loss\\
\hline
$\sigma$ & Std. dev of shadow fading\\
\hline
$X_c$ & Shadow fading correlation function\\
\hline
$\bm{S} \in\mathbb{R}^n, S_i$ at $i$ & Actual shadow fading measurements\\
\hline
$\bm{W} \in\mathbb{R}^n$ & Weights assigned to sensor measurements\\
\hline
$\bm{C_n} \in\mathbb{R}^{n\times n}$ & Covariance matrix of $\bm{S}$\\
\hline
$\bm{c_0} \in\mathbb{R}^n$ & cross-covariance of $\bm{S}$ and $S_0$ at point $P_0$\\
\hline
$\Sigma_i$ & RMS error of path loss estimation\\
\hline
\end{tabular}
\label{tab:notation}
\vspace{-0.2cm}
\end{table}
} 
\section{Reducing the Stochastic Method to Practice}

Examining the SM-0 approach suggests several ways to reduce the processing to practice. This is important because: (i) in reality, the median power\eat{$P_m$ (or bias in Kriging)} cannot be known precisely; (ii) it is very difficult to determine the correlations $c_{ij}$ and, thus, the correlation matrices $\bm{C_n}$ and $\bm{c_0}$; and (iii) even if it can be done, the real-time computation needed to obtain the weights for estimating $\mu_0$ can be quite high, especially as $n$ increases.

In this section, we present and analyze a more realizable approach, SM-1, which addresses the first concern about the median power\eat{$P_m$}; and following that, we present SM-2, which is slightly less ideal than SM-1 but addresses all three issues.% for making SM-0 practical.

\subsection{The First Method, SM-1}
\label{subsec:sm1}
For convenience, we begin by rewriting the power measured at the $i$-th sensor, where we assign $d_r = 1$ m, and we express $\Gamma$ as $10 \gamma$, where $\gamma$ is the path loss exponent. We assume, moreover, that $i = 1,..,n$ where $n$ is the number of sensors whose measurements are used to predict power at a particular unmeasured point (point 0). The power received at the $i$-th sensor is rewritten as
\begin{equation}
\label{eq:lo_PL}
\begin{aligned}
P_{r,i} &= A + 10\gamma\log d_i + S_i,\; i = 1,..,n,\\
&= (A + Z_n) + 10 \gamma \log d_i + (S_i - Z_n), 	
\end{aligned}	
\end{equation}
where
\begin{equation}
\label{eq:Z_n}
Z_n = \frac{1}{n} \sum_{i=1}^{n} S_i.
\end{equation}
i.e., $Z_n$ is the average of the $S$-values at the $n$ sensors. The reason for this reformulation will be made clear shortly.

A general approach to reducing the stochastic method to practice is as follows:
\begin{enumerate}
\item Use the $n$ measured values of $P_{r,i}$, along with least squares estimation (LSE), to estimate $A + Z_n$ and $\gamma$, leading to estimates, $A^{'}$ and $\gamma^{'}$, that are imperfect. % if $n$ is finite.
Due to the normal distribution of shadow fading values, the received power is also normally distributed with mean $P_m$, the median received power, and variance $\sigma^2$. In this case, LSE is equivalent to maximum likelihood estimation\cite{Osborne_leastsquares}.
\item Use the estimates $A^{'}$ and $\gamma^{'}$ to estimate the shadow fading term at each sensor, i.e.,
\begin{equation}
\label{eq:sAprx}
S^{'}_i = P_{r,i} - (A^{'} + 10 \gamma^{'} \log d_i ).
\end{equation}
\item To estimate $P_{r,0}$ at point 0, use the equation
\begin{equation}
\label{eq:PL0_est}
P^{'}_{r,0} = A^{'} + 10 \gamma^{'} \log d_0 + S^{'}_0.
\end{equation}
where $d_0$ is the distance from the emitter to point 0, and $S^{'}_0$ is a weighted sum over the $n$ estimates, $S_1^{'},S_2^{'},..,S_n^{'}$. The weights $w_1,w_2,..,w_n$ are the elements of $\bm{W}$, (\ref{eq:W}).
\end{enumerate}

What we call SM-1 is this three-step approach\eat{with optimal weights (see Section~\ref{subsec:est_weights})}, which requires knowing  the spatial correlation matrix of shadow fading; the second reduction method to be discussed later, SM-2, uses an ad hoc weighting approach that requires no such knowledge.

The reformulation of the power equation, (\ref{eq:lo_PL}), can now be explained: Conditioned on the $n$ values of $S_i$, the LSE algorithm seeks a solution $(A^{'}, \gamma^{'})$ that minimizes the sum over $i$ of
\begin{equation}
\label{eq:delta_Psq}
(\Delta P_{r,i})^2 = [P_{r,i} - (A^{'} + 10 \gamma^{'} \log d_i)]^2.
\end{equation}
In so doing, it implicitly stipulates that the ``noise" components in the $P_{r,i}$-values have a zero sum over $i$. Thus, it behaves as though the form of $P_{r,i}$ is as given in the bottom line of (\ref{eq:lo_PL}), where the term common to all $P_{r,i}$ is $(A + Z_n)$, hereafter referred to as $A^"$; and the ``noise" for each $P_{r,i}$-value is $(S_i - Z_n)$, hereafter referred to as $S_i^"$. (Note that the sum over $i$ of $S_i^"$ is zero.) The application of the LSE algorithm, therefore, yields $A^{'}$ as an approximation to $A^"$, and the $n$ $S_i^{'}$-values as approximations to the $S_i^"$-values.

The estimate for received power at point 0 is written as
\begin{equation}
\label{eq:PL0_est_1}
P^{'}_{r,0} = A^{'} + 10 \gamma^{'} \log d_0 + \sum_i w_i S^{'}_i.
\end{equation}
In view of the above, it can also be written as
\begin{equation}
\label{eq:PL0_est_2}
\begin{aligned}
P^{'}_{r,0} = A^{"} + 10 \gamma \log d_0 + \sum_i w_i S^{'}_i - \delta_{m,0},
\end{aligned}
\end{equation}
where, at any point $k$,
\begin{equation}
\label{eq:delm}
\begin{aligned}
\delta_{m,k} = \Delta A + 10 \Delta \gamma \log d_k,\\
\Delta A = A" - A^{'} \mbox{ and } \Delta \gamma = \gamma - \gamma^{'}.
\end{aligned}
\end{equation}
In Section \ref{app:deltaAgamma} of the Appendix, we derive errors $\Delta A$ and $\Delta \gamma$.

Using these results, we can write an expression for $(P_{r,0} - P_{r,0}^{'})$ which is the dB difference at point 0 between the estimated and actual received power. The result is
\begin{equation}
\label{eq:DeltaPL_0_1}
\begin{aligned}
\Delta P_{r,0}  &= P_{r,0} - P^{'}_{r,0}\\
&= \left(\delta_{m,0} - \sum_{i=1}^n w_i\delta_{m,i}\right) + \left(S^"_0 - \sum_{i=1}^n w_i S^{'}_i\right),
\end{aligned}
\end{equation}
where the first part reflects the total error caused by imperfect estimation of $A^"$ and $\gamma$, and the second part is the error due to imperfect estimation of $S_0^" = S_0  - Z_n$ \eat{unavailability of precise weights}. %(see Section \ref{app:SM1_weights} of the Appendix for more details).
While the formulation (\ref{eq:W}) has involved applying $\bm{W}$ to $S$-values, in actuality the path loss estimation uses $S^"$ (see \ref{eq:lo_PL}). Unfortunately, using the same formulation for the weights in terms of the correlation matrix of $S^"$ can occasionally lead to problems associated with an ill-conditioned $\bm{C_n}$ matrix depending on the geometry. Thus, to cope with this, one option is to calculate $\bm{W}$ using the correlation matrix of $S$, as in (\ref{eq:W}), but apply $\bm{W}$ to $S^"$.
This leads to very little degradation of the SM-1 results when compared with SM-0 results, as well as with results for other methods in the literature.

\eat{
While the formulation (\ref{eq:W}) has involved to apply $\bm{W}$ to $S$-values, in actuality received power estimation uses $S^"$ (see \ref{eq:lo_PL}). Unfortunately, for various sensor geometries, using the same formulation for the weights in terms of the correlation matrix of $S^"$ can occasionally lead to problems associated with an ill-conditioned (non-invertibility) $\bm{C_n}$. Thus, to cope with such cases, $\bm{W}$ using the correlation matrix of $S$, as in (\ref{eq:W}), has been applied to $S^"$. Such imperfect values of weights leads to additional performance degradation of the SM-1 when compared with the performance of SM-0.
}

Further solving for $\Delta P_{r,0}$, we have,
\begin{equation}
\label{eq:DeltaPL_0_2}
\begin{aligned}
\Delta P_{r,0} &= S_0 + \sum_{i=1}^n S_i\left(\alpha_i\log\left(\frac{d_0}{\prod_j d_j^{w_j}}\right) - w_i \right)\\
&+  \sum_{i=1}^n S_i\left(\left(1-\sum_j w_j\right) \left(\beta_i - \frac{1}{n}\right)  \right); j=\{1,..,n\}\\
\end{aligned}
\end{equation}
where
\begin{equation}
\label{eq:delm_2}
\begin{aligned}
&\alpha_i = \frac{\left(\sum_{j=1}^n \log d_j\right) - n\log d_i}
{n\sum_{j=1}^{n}\left(\log d_j\right)^2 - \left(\sum_{j=1}^n\log d_j\right)^2};\\
&\beta_i = \frac{\left(\log d_i\right)\left(\sum_{j=1}^n\log d_j\right) - (1/n)\left(\sum_{j=1}^n\log d_j\right)^2}
{n\sum_{j=1}^{n}\left(\log d_j\right)^2 - \left(\sum_{j=1}^n\log d_j\right)^2}.
\end{aligned}
\end{equation}
$\Delta P_{r,0}$ is seen to be a weighted linear sum of shadow fadings $S_1$ through $S_n$ at the sensors and $S_0$ at point 0. Therefore, $\Delta P_{r,0}$ is a zero-mean Gaussian random variable whose RMS value scales with $\sigma$.
%$\Delta P_{r,0}$ is a linear sum of shadow-fading $S_1$ through $S_n$ at sensors and $S_0$ at Point 0 with constant (distance) coefficients terms. Therefore, $\Delta P_{r,0}$ is quantified as a zero-mean Gaussian random variable with derivable RMS error.

\subsection{The Second Method: SM-2}
The obvious disadvantage of SM-1 is that the spatial correlation properties of shadow fading, and thus the weighting vector $\bm{W}$, are difficult to estimate in practice. However, the SM-1 analysis allows us to compute best-case bounds on attainable accuracy for any spatial correlation process. This provides a benchmark against which to compare less optimal but more practical schemes. Following the method of \cite{Zhao2004_Path}, where inverse distance weighting (IDW) is applied to the estimates $S_i$, we propose the nonparametric weighting function
\begin{equation}
\label{eq:W_sm2}
w_i = \frac{y_{0i}^{-\nu}}{\sum_{j=1}^n y_{0j}^{-\nu}}; i = 1,..,n,
\end{equation}
where $y_{0i}$ is the distance from sensor $i$ to point 0. We will use $\nu = 1$ in our calculations, as \cite{Zhao2004_Path} shows little variation in error performance for $\nu =$ 1, 2 or 3. Also note that, for any choice of $\nu$, the $n$ weights add to 1, as in \cite{Zhao2004_Path}, which is not necessarily true for the weights used in SM-1.

\eat{
The obvious disadvantage of SM-1 is that the precise correlation, and thus the weight vector $\bm{W}$ will not be available in practice. However, SM-1 analysis allows us to compute best-case bounds on attainable accuracy for any specified spatial correlation function. This provides a benchmark against which to compare the less optimal but more practical weight-selection approach SM-2 proposed in our previous work in \cite{Zhao2004_Path}.

The weights assigned to the sensor $i$ for estimation of $S_0^"$ is given by
\begin{equation}
\label{eq:W_sm2}
w_i = \frac{d_{0i}^{-\nu}}{\sum_{j=1}^n d_{0j}^{-\nu}}; i = 1,..,n,
\end{equation}
where $d_{0i}$ is the distance sensor $i$ and the point $0$ under estimation. We use $\nu = 1$ as the study \cite{Zhao2004_Path} showed that the difference in error performance is small across different choices of $\nu$.

Furthermore, irrespective of choice of $\nu$, the summation $\sum_i w_i = 1$ is over all sensors. This is a desirable characteristic as the error due to the unknown value of $Z_n$ would not appear in the expression $\mu_0^"$, derived from \ref{eq:mu5} and \ref{eq:lo_PL} as
\begin{equation}
\label{eq:s_55_est}
\begin{aligned}
\mu_0^{"} = \sum_{i=1}^n w_i S_i - Z_n
		  = \sum_{i=1}^n w_i S_i^" + \left(\sum_{i=1}^n w_i - 1\right)Z_n.
\end{aligned}
\end{equation}
In case of SM-1, such equality of weights summation to 1 is not guaranteed. We have discussed weights summation aspects more in the Appendix \ref{app:SM1_weights}.

Fig.~\ref{fig:weight_dist} shows weights assigned to each of sensors with respect to its distance from the Point 0 for particular sensor geometry and environmental parameters. As shown, the assignments of weights for SM-2 closely follows the assignment for SM-1 but using easily available information (distance measurements only). We note that this trend essentially remains the same for other environmental parameters with very small to no loss of accuracy (results are provided in the next Section).
\begin{figure}[t]
\begin{center}
\includegraphics[width=2.25in]{pics/weight_dist_E-100_0_sig5_D_Xc1.eps}
%\vspace{-1em}
\caption{Weights as a function of distance for SM-1 and SM-2 (Particular scenario with sensor forms a square geometry with $n = 4$ and side of a square = 640 m, $\sigma = 5$ dB, and $X_c$ = 640 m)}
\label{fig:weight_dist}
\end{center}
\vspace{-1.5em}
\end{figure}
}

\eat{
\section{Path loss based interpolation}
\label{sec:SM1}
In this section, we propose two path loss based interpolation approaches- \textit{SM-1} and \textit{SM-2}. Both the approaches use a Least-Square estimation (LSE) to compute median path loss at any given arbitrary point. To compute shadowing, SM-1 and SM-2 employ different weighting-based approaches as described later in the section. To consider shadow fading correlation among $n$ sensors and an arbitrary point, we have rearranged path loss model mathematically. We begin with the explanation this recasted \textit{local path loss model} as next-

\subsection{Path Loss Estimation Algorithm}
We propose the following procedure for path loss estimation between transmitter $E$ and any arbitrary point $P_0$ in the area bounded by a given set of sensors.
\begin{enumerate}
\item \textbf{Given/known parameters}\\
For given $n$ sensors, path loss vector $\bm{PL}\in \mathbb{R}^n$ and corresponding distance vector $\bm{d}\in \mathbb{R}^n$ with distances between $E$ and sensors are available.

\item \textbf{Estimation of $A^"$ and $\gamma$}\\
Apply Least-Square Estimation (LSE) to estimate values $A^"$ and $\gamma$ in Eq.~\ref{eq:lo_PL} by assuming functional approximations as
\begin{equation}
\label{eq:funAprx}
\begin{aligned}
\bm{PL}^{'} = A^{'} + 10\gamma^{'}\log_{10}\bm{d}
\end{aligned}
\end{equation}
where $\bm{PL}^{'}$ is estimated path loss vector; and estimated values $A^{'}$ and $\gamma^{'}$ minimize the squared distance of $\|\bm{PL} - \bm{PL}^{'}\|$.

Furthermore, while estimating $A^"$ and $\gamma$, LSE combines common bias $Z_n$ along with the median path loss $P_m$. Thus, consideration of local path loss model, given by Eq.~\ref{eq:lo_PL}, is justified.

\item \textbf{Estimation of shadow fading at sensors}\\
Calculate $\bm{S^{'}} \in \mathbb{R}^n$ at sensors, the estimates of $\bm{S^{"}}=\bm{S}-Z_n$, as the differences between measured path loss and estimated median path loss, as follows:
\begin{equation}
\label{eq:sAprx}
S^{'}_i = \bm{PL} - \bm{PL}^{'}.
\end{equation}
\item \textbf{Estimation of median path loss at $\bm{P_0}$}\\
Estimate the median path loss $PL^{'}_{m_0}$ at $P_0$ as
\begin{equation}
\label{eq:EstMedPL}
PL^{'}_{m,0} = A^{'} - 10\gamma^{'}\log_{10}d_0.
\end{equation}
\item \textbf{Estimation of shadow fading at $\bm{P_0}$}\\
Shadow fading $S_0^"$ at $P_0$ is estimated as a weighted sum of shadow fading values at sensors, i.e., the estimate of $S_0^"$ is
\begin{equation}
\label{eq:s5}
S^{'}_0 = \bm{W^T} \bm{s^{'}}
\end{equation}
with the weight vector $\bm{W} = [w_1\;w_2\;\ldots\; w_n]^T$. Note that the estimate of $S_0^"$ is called $S^{'}_0$. The procedures for calculating the weights $\bm{W}$ for SM-1 and SM-2 are given in the following section.

\item\textbf{Estimation of path loss at $\bm{P_0}$}\\
From equations \ref{eq:EstMedPL} and \ref{eq:s5}, we have
\begin{equation}
\label{eq:PL0_est}
PL^{'}_{0} = PL^{'}_{m,0} + S^{'}_0.
\end{equation}

\end{enumerate}

In above algorithm in step (2), linear regression is implemented using Least-square estimation (LSE). Due to normal distribution of shadow fading values, path loss is in fact normally distributed with $\mathbb{G}(PL_m,\sigma^2)$, where $PL_m$ is the median path loss. Thus, in this case, LSE is equivalent to maximum likelihood estimation\cite{Osborne_leastsquares}.

\subsection{SM-1}
In our first proposed approach- SM-1, weights are based on stochastic analysis as given by Eq.~\ref{eq:W}
\[\bm{W} = \bm{(c_0^T C_n^{-1})^T}.\]
which are functions of correlation matrix of $\bm{S}/S_0$. But the path loss estimation algorithm (steps 1-6) uses measurements of $\bm{S^"}$ and calculates $S^"_0$. Thus, ideally, $W$ should be calculated using covariance matrices of $\bm{S^"}/S^"_0$.
For defined local path loss model (see Eq.~\ref{eq:lo_PL}), $\bm{W}$ is not derivable. As an example shown in Fig.~\ref{fig:sq_geo}, we consider a square geometry with $n = 4$ sensors. For each sensor, we have
\[s^"_i = s_i - \frac{s_1+s_2+s_3+s_4}{4}; i = 1,..,4.\]
\begin{figure}[t]
\begin{center}
\includegraphics[height=1.25in,width=1.25in]{pics/sq_geo.eps}
%\vspace{-1em}
\caption{An example of sensors geometry (square) for which covaraiance matrix for $\bm{S"}$ does not exits.}
\label{fig:sq_geo}
\end{center}
\vspace{-1.5em}
\end{figure}
For correlation function given by Eq.~\ref{eq:auto-co}, each element of covariance matrix $\bm{C_4^"}$ of $\bm{s^"}$ is given as
\begin{equation}
\label{eq:corr_s"}
\bm{C_4^"}(i,j) = \left\{ \begin{array}{ll}
\frac{1}{4}(3a - 2ab - ac) & \mbox{if $i = j$};\\
\frac{1}{4}(-a + 2ab - ac) & \mbox{if $\|i - j\| = D$};\\
\frac{1}{4}(-a - 2ab + 3ac) & \mbox{if $\|i - j\| = \sqrt{2}D$},
\end{array}
\right.
\end{equation}
where
\[\begin{aligned}
a = \sigma^2,& b = \exp\left(-\frac{D}{X_c}\right),& c = \exp\left(-\frac{\sqrt{2}D}{X_c}\right)
\end{aligned}\]
and $i, j \in \{1,..,4\}$. The determinant of such $\bm{C_4^"}$ is zero. Thus, $\bm{C_4^"}$ is not a positive semi-definite and real value of $(\bm{C_4^"})^{-1}$, thus $\bm{W}$ using $\bm{C_4^"}$, does not exist. This mathematical explanation can be extended to other sensor geometries and number of sensors.  Therefore, we follow $\bm{W}$ derived by Eq.~\ref{eq:W} for vector $\bm{S^"}$ as well.

From Eq.~\ref{eq:mu5} and \ref{eq:lo_PL}, the mean value of $s_0^{"}$ is
\begin{equation}
\label{eq:s_55}
\mu_0^{"} = \bm{W^TS} - Z_n.
\end{equation}
which is recasted as
\begin{equation}
\label{eq:s_55_est}
\mu_0^{"} = \sum_{i=1}^n w_i\bm{S}_i^" + \left(\sum_{i=1}^n w_i - 1\right)Z_n.
\end{equation}
Now the first term is a weighted sum over measured $\bm{S^"}$ values, but the second term still involves the unmeasured term $Z_n$. This problem goes away if the weights sum to 1. We also note that the summation of weights depends on the number and geometry of sensors involved in the measurement. We will elaborate on this later in Section~\ref{sec:eval}.

\subsection{SM-2}
The obvious disadvantage of SM-1 is that the vector $\bm{W}$ will not be available in practice. However, SM-1 analysis allows us to compute best-case bounds on attainable accuracy for any specified spatial correlation function. This provides a benchmark against which to compare less optimal but more practical weight-selection approaches SM-2 as proposed in our previous work in \cite{Zhao2004_Path}.

The weights assigned to the sensor $i$ for estimation of $S_0^"$ is given by
\begin{equation}
\label{eq:W_sm2}
w_i = \frac{d_{0i}^{-\nu}}{\sum_{j=1}^n d_{0j}^{-\nu}}; i = 1,..,n,
\end{equation}
where $d_{0i}$ is the distance sensor $i$ and the point $P_0$ under estimation. We use $\nu = 1$ as the study \cite{Zhao2004_Path} show that difference in error performance is small for various choices of $\nu$. Also, irrespective of choice of $\nu$, the summation $\sum_i w_i = 1$ over all sensors. This is a desirable characteristics as the error due to unknowing value of $Z_n$ would not appear as given by Eq.~\ref{eq:s_55_est}.
} 
\section{Evaluation}
\label{sec:comp}

\subsection{Methodology}
\label{subsec:setup}
\begin{figure}[t]
\begin{center}
\includegraphics[width=2.6in]{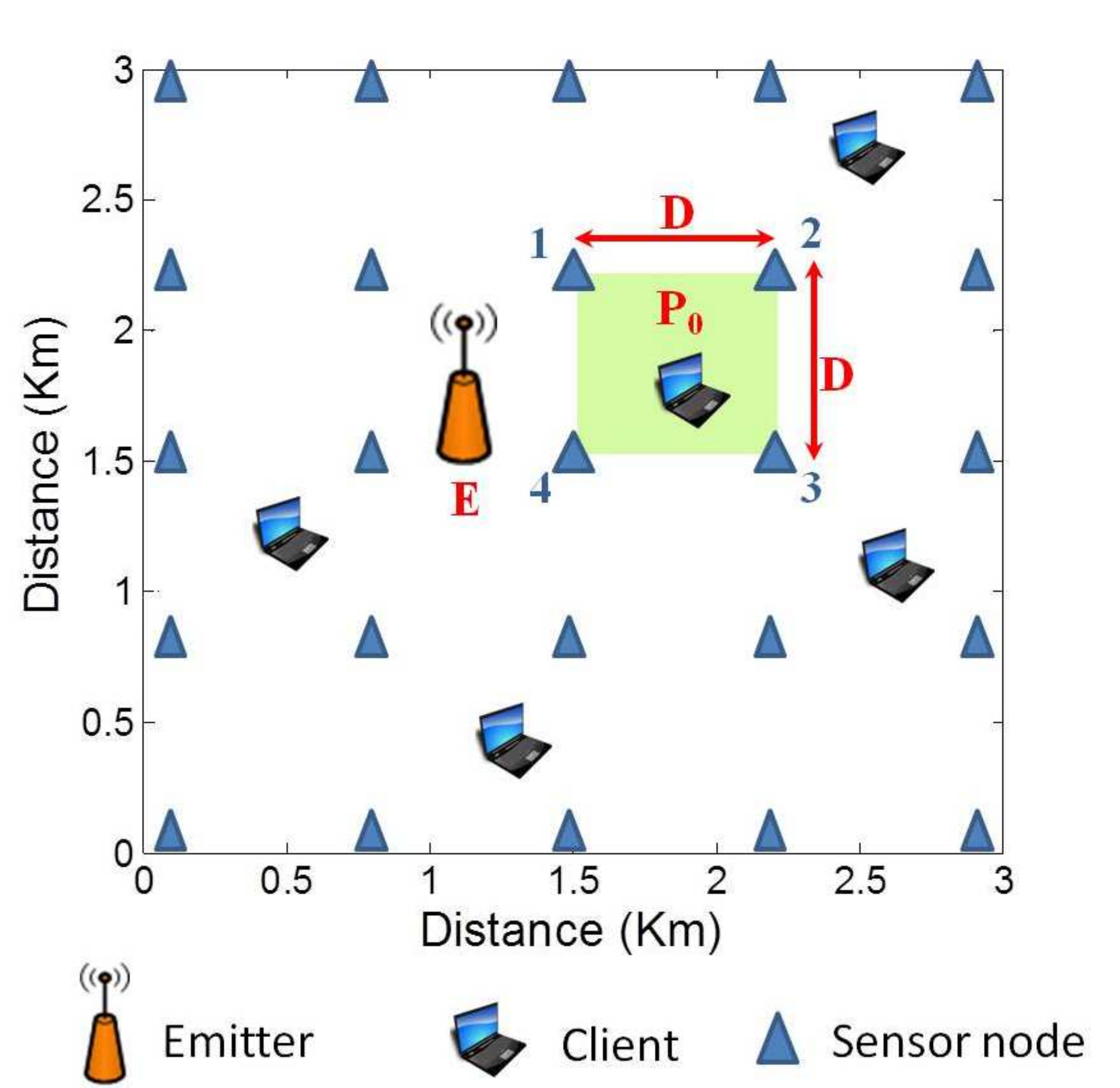}
%\vspace{-1em}
\caption{Problem geometry to be studied. The sensor layout defines a square grid, with each square having side $D$. Computations are made for the square shown shaded, with sensors 1, 2, 3 and 4. Interpolation methods are applied to points inside the square to estimate received power from the emitter at location $E$.}
\label{fig:senGrid_tx}
\end{center}
\vspace{-0.5em}
\end{figure}
For the sake of concreteness, we postulate a particular geometry, as shown in Fig.~\ref{fig:senGrid_tx}. In a 3-km x 3-km area, a  grid of sensors are superimposed where the sensors are separated by distance $D = 640$ m. The postulated geometry is a typical coverage area that might be used, e.g., in a cellular network with a primary emitter (base station), primary clients and secondary emitters. We will use this geometry to quantify the accuracy of specific approaches. The proposed approaches can be scaled to other dimensions and extended to other geometries\footnote{Another possibility is tessellating hexagons in place of squares, as in studies of cellular networks}. Towards the objective of building a radio map, we focus on one of the sensor squares, shown by the shaded region in Fig.~\ref{fig:senGrid_tx}, and can apply our methodology to each square within the large grid. We will compute the RMS interpolation error at points inside the square by averaging over the statistical ensemble of shadow fading; we can also regard this as spatial averaging over all the squares in the grid, assuming the propagation model is statistically stationary.
We assume that there is an emitter $E$ external to this square area and all given sensors (here, $n=4$) scans the same band, where each of the sensors measures and reports the received power from $E$.
We assume the sensor measurements are sufficiently wideband that the effects of local multipath fading are averaged out. We continue to assume that the emitter location is known; later, however, we discuss how the case of unknown location and/or multiple emitters might be handled.

\begin{table}[t]
\caption{Simulation parameters}
\centering
\begin{tabular}{| l | l |}
\hline
\textbf{Parameter} & \textbf{Value}\\
\hline
Path loss model & $15.3 + 10(3.76)\log_{10}d$\\
(3GPP suggested model\cite{TR3GPP}) & $A= 15.3, \gamma = 3.76$\\
\hline
Shadow fading spread & $\sigma=5$ dB\\
\hline
Number of sensors & $n = 4$\\
\hline
Length of a side of the square & $D = 640$ m\\
\hline
Assumed coordinates for $n$ sensor & $(0,0), (0,640), (640,640), (640,0)$\\
\hline
\end{tabular}
\label{tab:simu_para}
\vspace{-0.2cm}
\end{table}

\begin{figure}[t]
\centering
\subfigure[Comparisons of SM-0, SM-1, SM-2]{
	\includegraphics[width=2.25in]{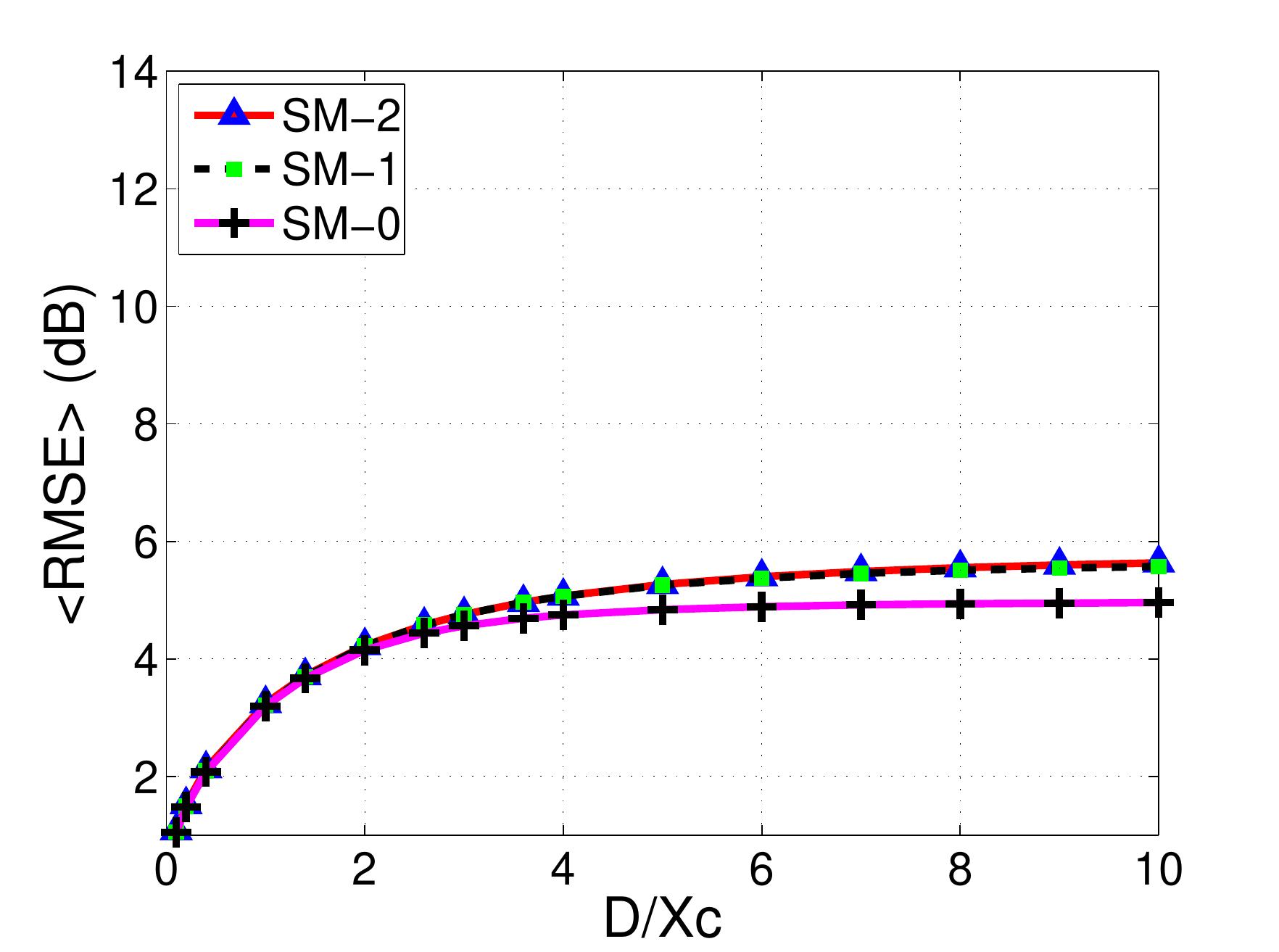}
    \label{fig:SM012_circular_E-100_0}
	\vspace{-0.6cm}
}
\subfigure[Comparisons of SM-2 with Nearest Neighbor (NN), Natural Neighbor (NaN), Inverse Distance Weights (IDW)]{
	\includegraphics[width=2.25in]{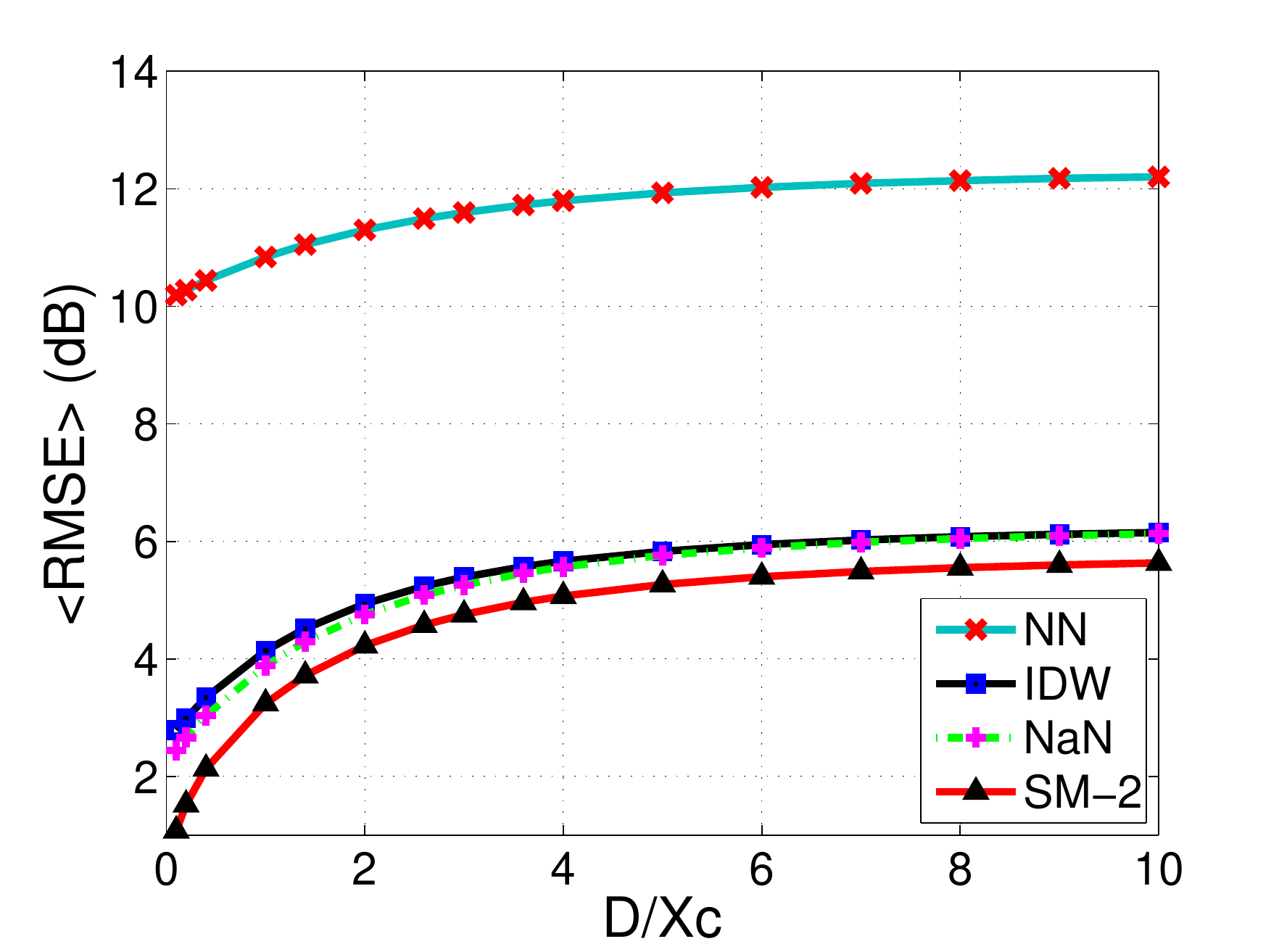}
    \label{fig:comAll_circular_E-100_0}
	\vspace{-0.6cm}
}
\caption[Optional caption for list of figures]{Spatial averages over the square of RMS interpolation error, for different methods, as a function of $D/X_c$. Exponential correlation function, $\sigma = 5$ dB and Emitter at $E(-100,0)$. 
%%\subref{fig:NvswoACAwACA}, \subref{fig:NvsACAgainRanFix}
}
\label{fig:RMSE_E-100_0}
\vspace{-0.5cm}
\end{figure}

\begin{figure}[t]
\centering
\subfigure[Probability distribution function]{
	\includegraphics[width=2.25in]{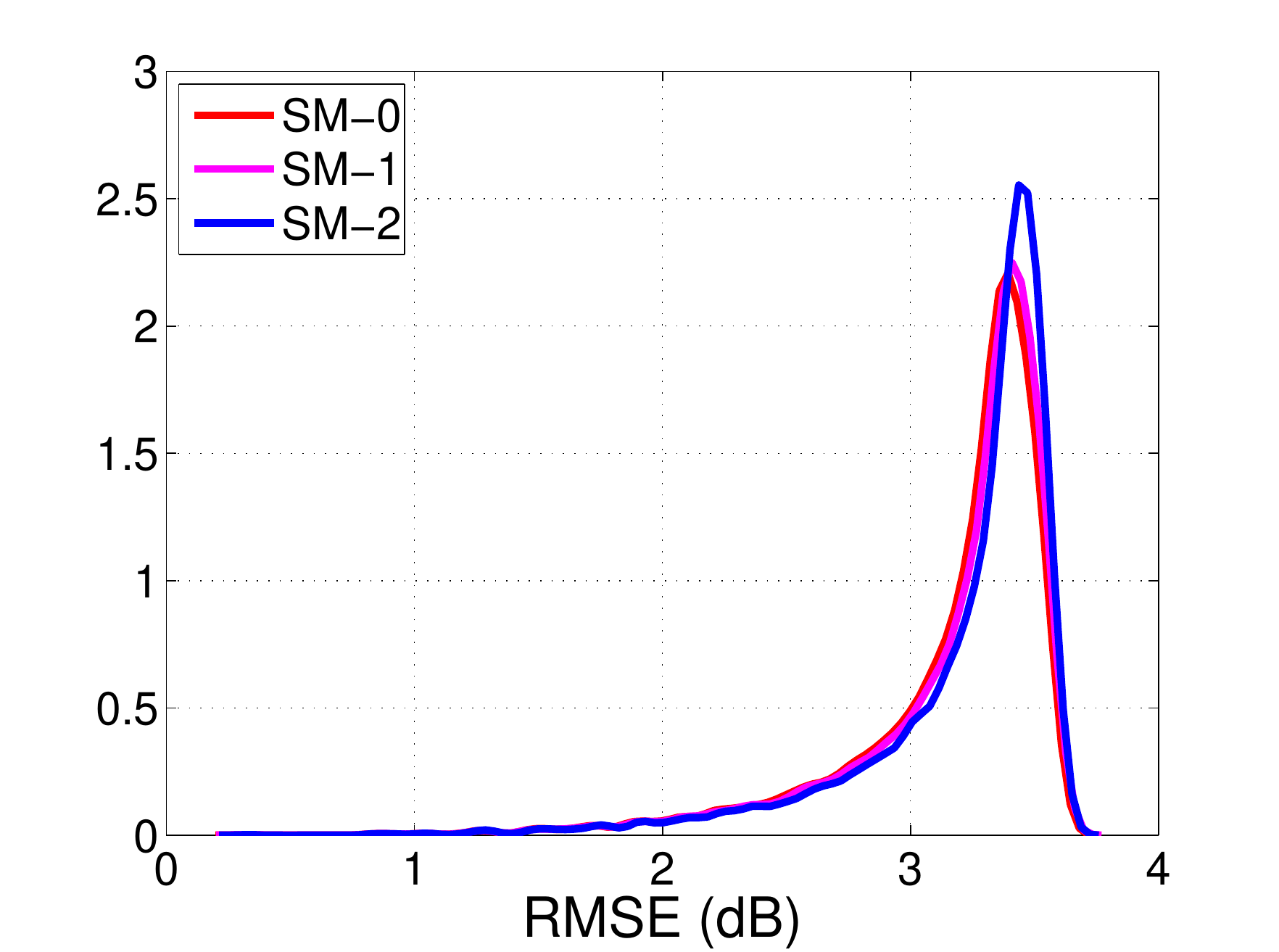}
    \label{fig:SM012_circular_E-100_0_rmsePDF}
	\vspace{-0.6cm}
}
\subfigure[Cumulative distribution function]{
	\includegraphics[width=2.25in]{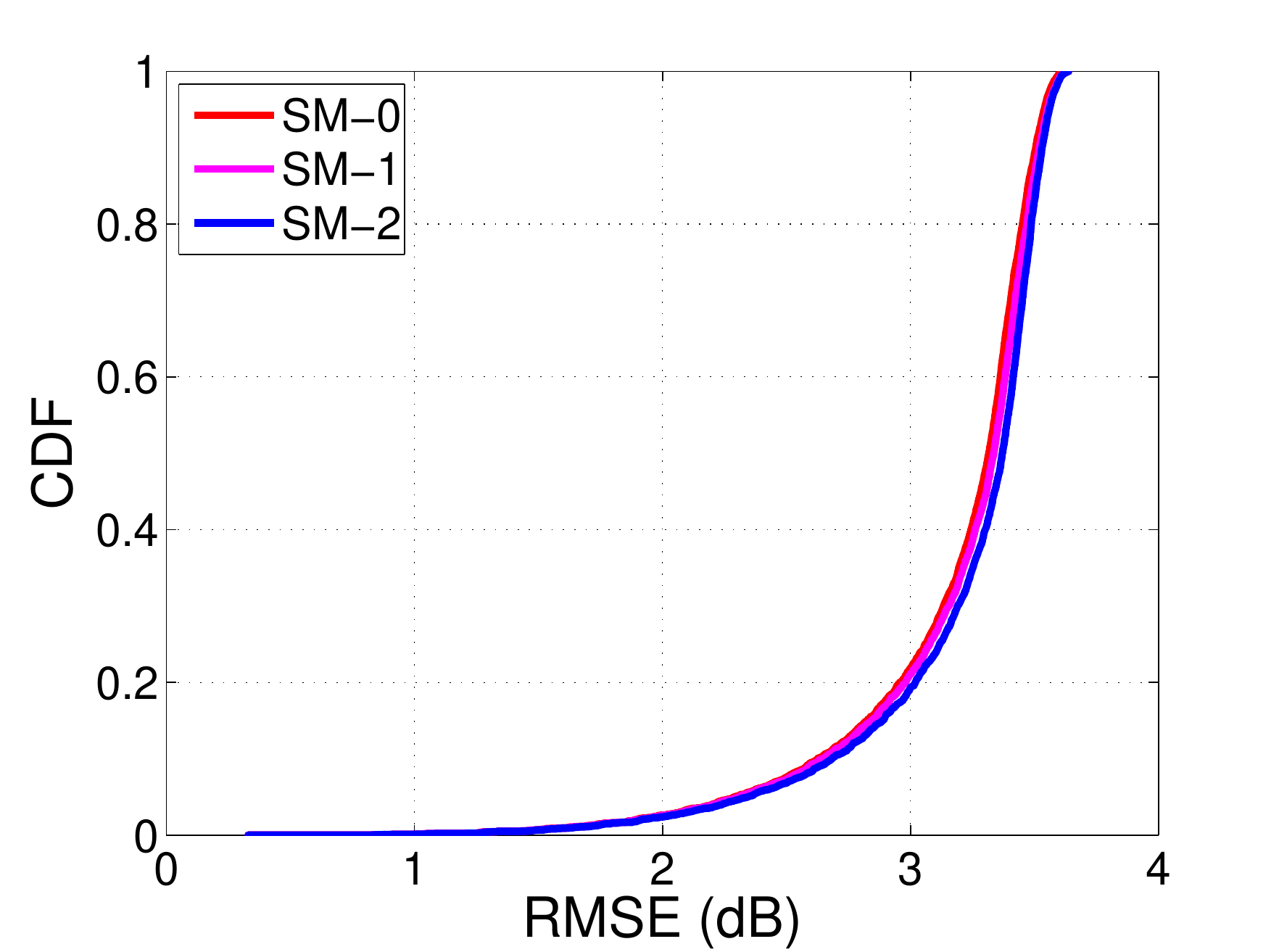}
    \label{fig:SM012_circular_E-100_0_rmseCDF}
	\vspace{-0.6cm}
}
\caption[Optional caption for list of figures]{Spatial distributions of RMS interpolation error over the square, for the SM family and $D/X_c = 1$. Exponential correlation function, $\sigma = 5$ dB and Emitter at $E(-100,0)$.
%%\subref{fig:NvswoACAwACA}, \subref{fig:NvsACAgainRanFix}
}
\label{fig:RMSE_E-100_0_CDFPDF}
\vspace{-0.5cm}
\end{figure}

Here the impact of emitter $E$ at any arbitrary point $0$ is determined by collecting measurements at the $n=4$ sensors surrounding point $0$ and applying the proposed interpolation algorithms - SM-0, SM-1, and SM-2. We evaluate the performance of each algorithm with respect to RMS error as described in previous sections. Further, we provide a comparison of our proposed algorithms with several common interpolation techniques: Nearest Neighbor (NN), Inverse Distance Weighting (IDW), and Natural Neighbor (NaN). NN is the simplest interpolation technique where estimation at the point $0$ is equal to the measurement at the sensor nearest to point $0$. Both IDW and NaN provide estimates that are weighted sums of the $n$ sensor measurements, with the sum of the weights being 1. For IDW, each weight is based on distance to the sensor; for NaN, the weights are based on areas, using Voronoi cells\cite{Ureten2012_comparisonCartography}.

Point 0 can be anywhere inside the square, and we will find the RMS interpolation error at many such points, specifically, points on a 64 $\times$ 64 array distributed uniformly over the square. At every one of the 64 $\times$ 64 = 4096 points, we will obtain the RMS interpolation error by averaging over 10000 realizations (or ‘instances’) of $[S_0, S_1, S_2, S_3, S_4]$; and then we will average over the square. The overall spatial average will be represented by the metric $<$\textit{RMSE}$>$, 
\eat{
In a simulation based comparison study, the RMS error was computed at $64 \times 64$ arbitrary points in a given square. At each arbitrary point 0, instances of $[S_0, S_1, S_2, S_3, S_4]$ are generated from the joint Gaussian distribution of the $S$-values, for which true received power is known.
%$10000$ instances of random received power (represents the true received power) were generated.
For each of the interpolation methods, the received power at each point 0 is estimated, followed by the computation of the RMS error, which was averaged over the number of instances. The overall RMS error for the square with $m = 4096$ arbitrary points is represented by a single value
}
\begin{equation}
<\mbox{\textit{RMSE}}> = \sqrt{\frac{\sum_{j=1}^m x_j^2}{m}},
\end{equation}
where $<  >$ in this case denotes a spatial average, and $x_j$ is the RMS error at the $j$-th point in the $64 \times 64$ array. Other simulation parameters are listed in Table~\ref{tab:simu_para}.

\begin{figure*}[t]
\centering
\subfigure[RMSE for SM-0, SM-1, SM-2 when $E(-100,320)$]{
	\includegraphics[width=2.25in]{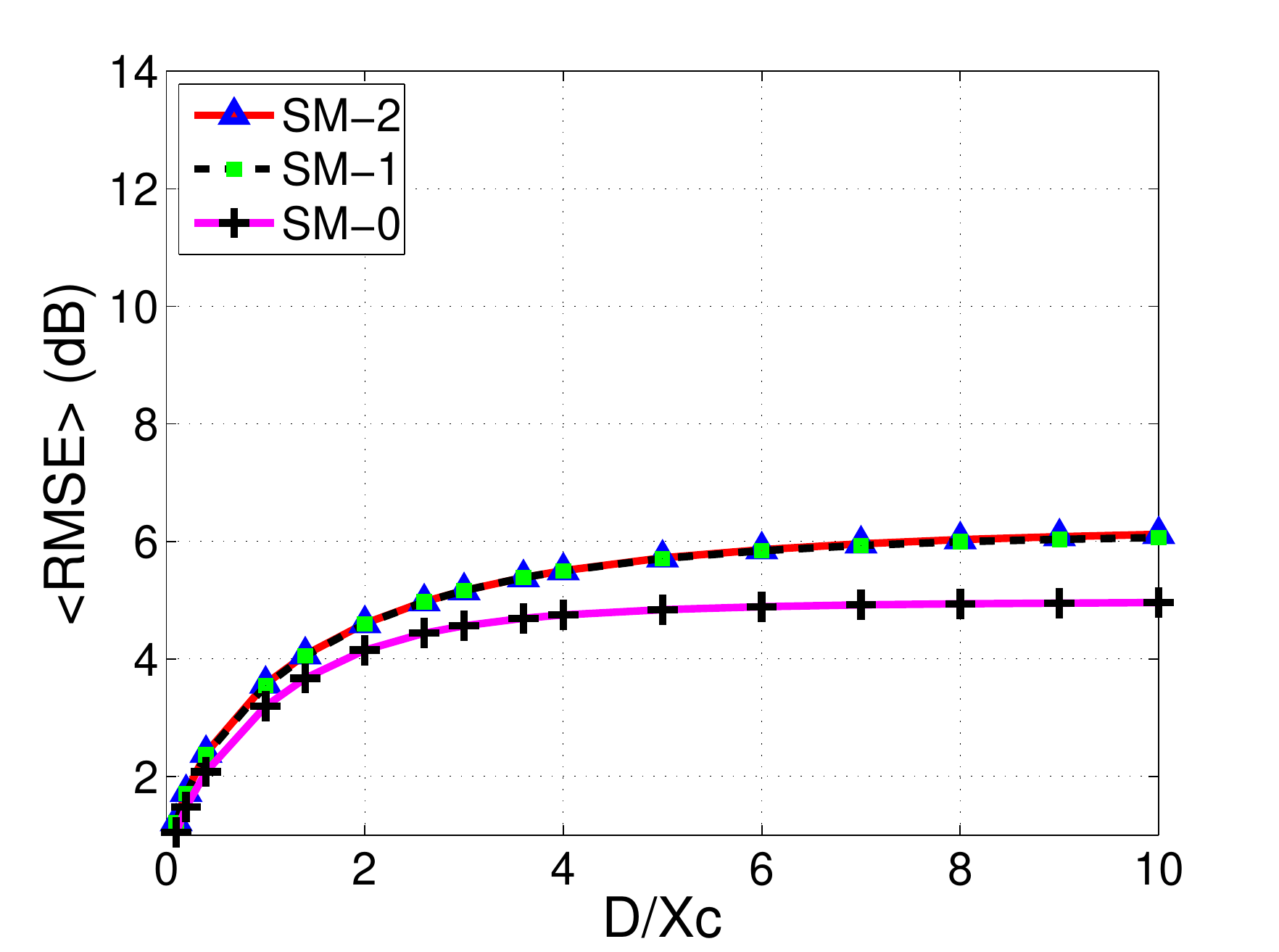}
    \label{fig:SM012_circular_E-100_320}
	\vspace{-0.6cm}
}
\subfigure[RMSE comparison of SM-2 with NN, NaN, IDW $E(-100,320)$]{
	\includegraphics[width=2.25in]{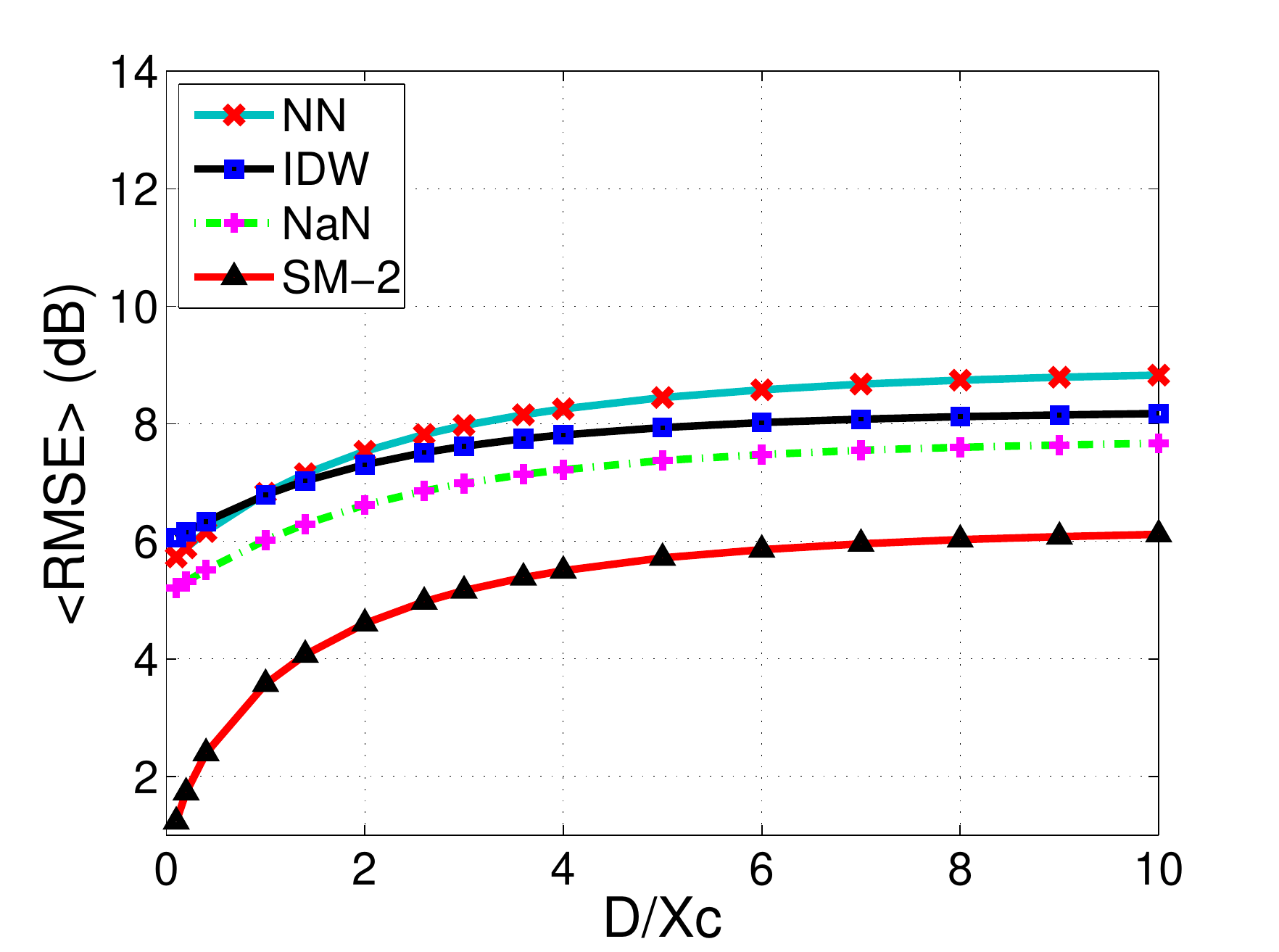}
    \label{fig:comAll_circular_E-100_320}
	\vspace{-0.6cm}
}
\subfigure[RMSE for SM-0, SM-1, SM-2 when $E(-400,-400)$]{
	\includegraphics[width=2.25in]{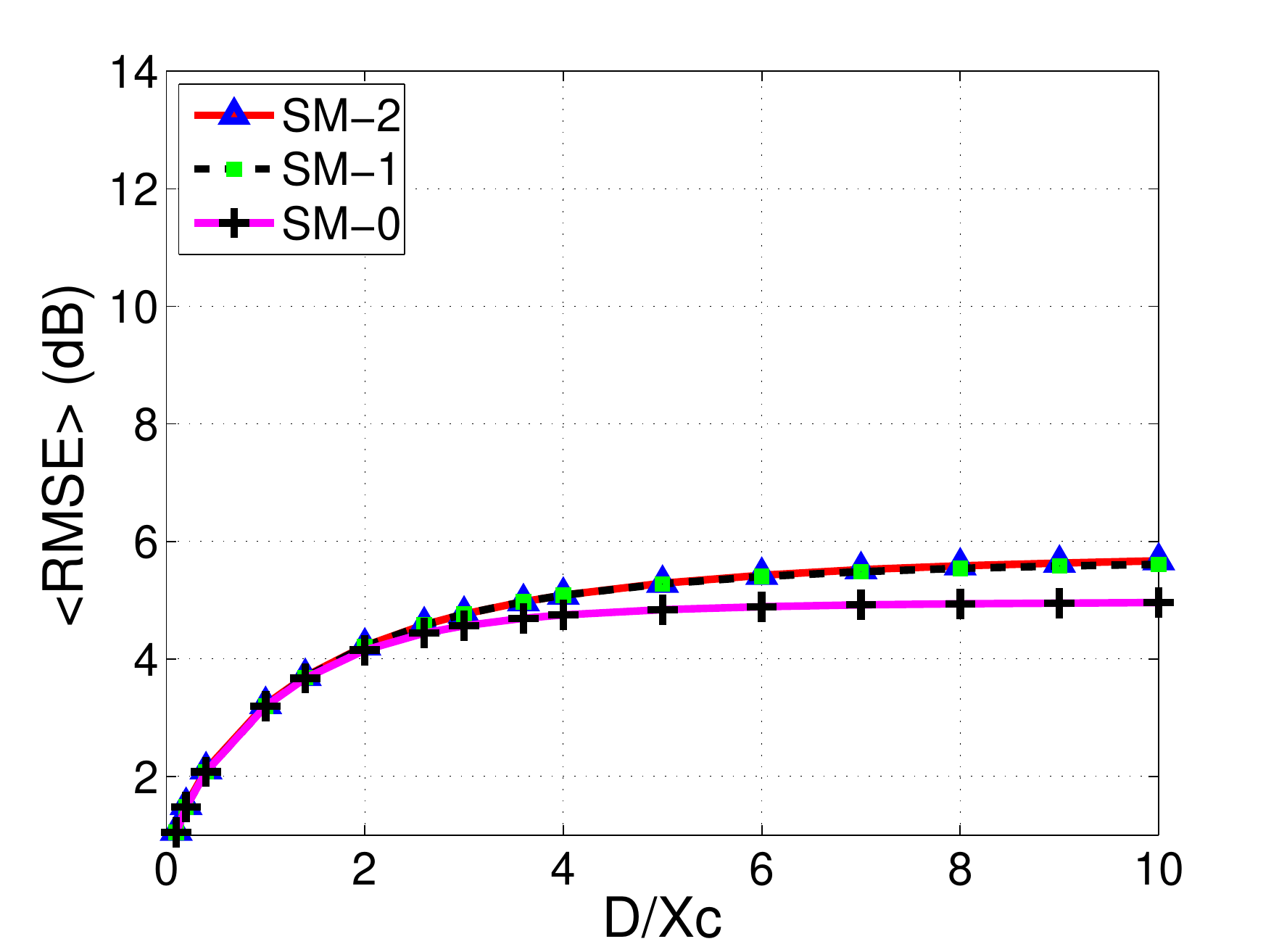}
    \label{fig:SM012_circular_E-400_400}
	\vspace{-0.6cm}
}
\subfigure[RMSE comparison of SM-2 with NN, NaN, IDW $E(-400,-400)$]{
	\includegraphics[width=2.25in]{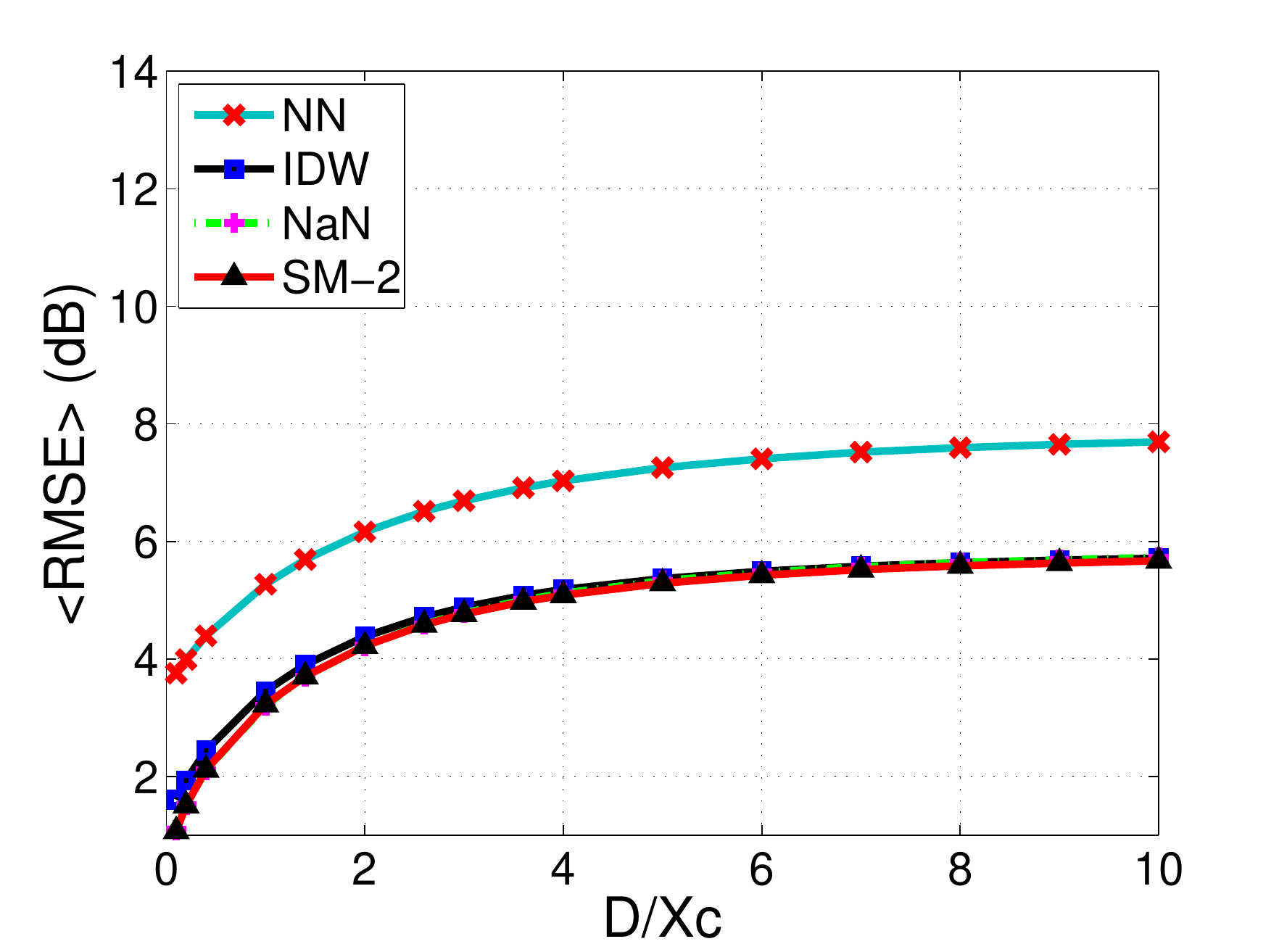}
    \label{fig:comAll_circular_E-400_400}
	\vspace{-0.6cm}
}
\caption[Optional caption for list of figures]{Effect of emitter location on $<$\textit{RMSE}$>$ vs. $D/X_c$. Exponential correlation function and $\sigma = 5$ dB.%%\subref{fig:NvswoACAwACA}, \subref{fig:NvsACAgainRanFix}
}
\label{fig:RMSE_emitter}
\vspace{-0.5cm}
\end{figure*}

\subsection{Effect of Sensor Spacing}
The most important parameter of the sensor network design is the sensor density, e.g., the number of sensors per unit area. This can also be captured by the nominal spacing between sensors which, in our problem geometry, Fig.~\ref{fig:senGrid_tx}, corresponds to the side of the square, $D$.

Moreover, the impact of the spacing depends, not on its absolute value, but on that value relative to the distance over which shadow fading decorrelates. From (\ref{eq:auto-co}) (or related correlation functions), we can use the correlation distance $X_c$ for this purpose, and examine $<$\textit{RMSE}$>$ as a function of $D/X_c$. Typically, $X_c$ varies from several meters to a few hundred meters, depending on the type of terrain \cite{Goldsmith1994_Error,Ureten2012_comparisonCartography}.

\eat{
The interpolation algorithm performance is characterized against the normalized sensor density $D/X_c$ where $D$ is the length of a side of the square, and $X_c$ is the shadow fading correlation distance. Typically, $X_c$ varies between a few meters to a few hundred meters\cite{Goldsmith1994_Error,Ureten2012_comparisonCartography}. Our analysis shows that the performance of the proposed approach \textit{SM} remains the same for a fixed ratio $D/X_c$ irrespective of the individual values of $D$ and $X_c$.
}

Fig.~\ref{fig:RMSE_E-100_0} shows $<$\textit{RMSE}$>$ as a function of $D/X_c$ when the emitter is located at $E(-100, 0)$ with respect to given sensor coordinates (see Table~\ref{tab:simu_para}). Fig.~\ref{fig:SM012_circular_E-100_0} compares $<$\textit{RMSE}$>$ for proposed approaches SM-0, SM-1, and SM-2. As expected, SM-0 provides the lower bound of the estimation error. We note that error curves of SM-1 overlaps with SM-2, even though SM-2 lacks knowledge of the spatial correlation function. Furthermore, Fig.~\ref{fig:comAll_circular_E-100_0} shows that SM-2 estimates received power with the lowest RMS error when compared with the NN, IDW and NaN methods.

To affirm that $<$\textit{RMSE}$>$ is an appropriate metric, we computed the probability density function (pdf) and its associated cumulative distribution function (CDF) over the $m =$ 4096 array points within the square. Results are shown in Fig. \ref{fig:RMSE_E-100_0_CDFPDF} for the particular case $D/X_c  =$ 1. The major finding is the same for other values of $D/X_c$ as well, namely, that the RMS error is fairly uniform over the square. From the figure, for example, we observe for $D/X_c  =$ 1 and the three SM approaches, that the RMS error is within 0.3 dB of $<$\textit{RMSE}$>$ at 80$\%$ of the points in the array.

\subsection{Effect of Emitter Location}
For the family of SM approaches, the location of the emitter does not affect the estimate of $S_0$ but can affect the accuracy in estimating the median, $P_m$, at a given point. To demonstrate this impact, Fig.~\ref{fig:RMSE_emitter} shows plots of $<$\textit{RMSE}$>$ vs. $D/X_c$ for two distinct locations. For the three SM approaches, the impact of emitter location is seen to be relatively small (see \ref{fig:SM012_circular_E-100_320} and \ref{fig:SM012_circular_E-400_400}); comparing \ref{fig:comAll_circular_E-100_320} and \ref{fig:comAll_circular_E-400_400}, we see that the impact on NN, NaN and IDW is greater. These trends are evident across a range of emitter locations.

\eat{The family of \textit{SM} approaches takes distances between sensors, emitters and arbitrary points as an input for the estimation of received power. Thus, the sensors-to-emitter geometry influences the accuracy of the estimation algorithms. To illustrate this behavior, we study the performance of the RMS error for two emitter locations- $E(-100, 320)$ and $E(-400,-400)$.
\eat{
Both of these emitter locations cause degeneracy, that is, multiple sensors have the same emitter-sensor distance $d$. This leads to inaccuracies in the LSE while estimating path loss parameters as given in Section \ref{subsec:sm1}. The effect of degeneracy for \textit{SM} is reflected by an increase in $<$\textit{RMSE}$>$ of up to 1 dB as shown in Fig.~\ref{fig:RMSE_emitter}.
}
Also, we note that the variation around $<$\textit{RMSE}$>$ for \textit{SM} is small compared to that for NN, IDW, and NaN when the emitter location is varied. Thus, from an overall perspective, we may consider \textit{SM} to be insensitive to emitter location.
}

\subsection{Effect of Correlation Function}
The shape of the spatial correlation function in (\ref{eq:auto-co}) is exponential, but other shapes may prevail, depending on the topography. To show the robustness of RMS error results to this shape, we now consider two other cases$-$
\subsubsection{Gaussian correlation function}
We now consider the correlation function\cite{abrahamsen1997review}
\begin{equation}
\label{eq:auto-co-Gauss}
c_{ab} = \sigma^2 \exp\left[-\left(\frac{d_{ab}}{X_c}\right)^2\right]
\end{equation}
Comparing Figs. \ref{fig:SM012_gauss_E-100_0} and \ref{fig:comAll_gauss_E-100_0} with Figs. \ref{fig:SM012_circular_E-100_0} and \ref{fig:comAll_circular_E-100_0}, we see no substantial difference between the exponential and Gaussian correlation functions, (\ref{eq:auto-co}) and (\ref{eq:auto-co-Gauss}).These functions can be considered circular, i.e., in each case, the locus of constant correlation is a circle. We next consider a correlation function that depends on direction as well as distance separation.
\subsubsection{Elliptical correlation function}
\begin{figure*}[t]
\centering
\subfigure[RMSE for SM-0, SM-1, SM-2 for Gaussian correlation function]{
	\includegraphics[width=2.25in]{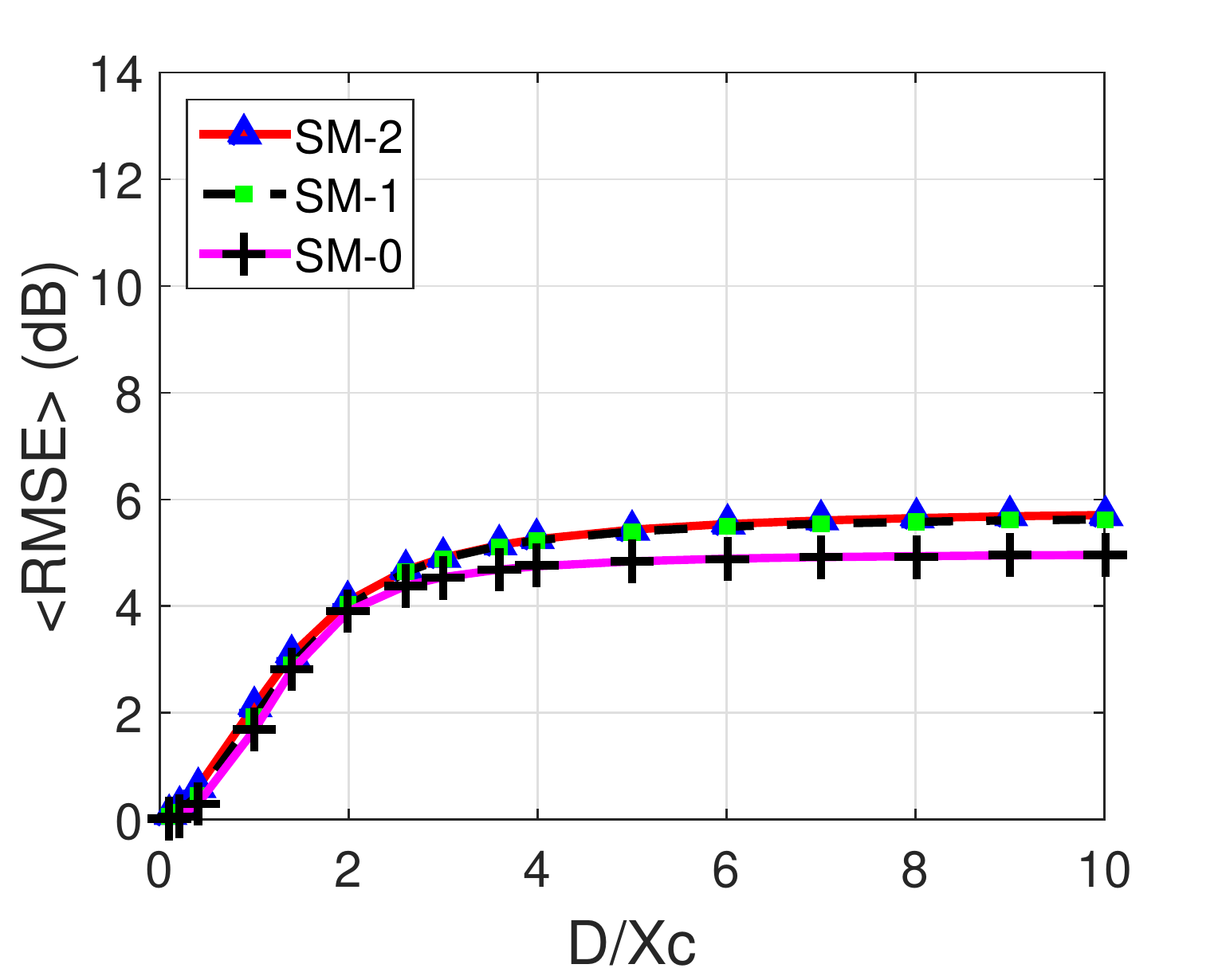}
    \label{fig:SM012_gauss_E-100_0}
	\vspace{-0.6cm}
}
\subfigure[RMSE comparison of SM-2 with NaN, IDW for Gaussian correlation function]{
	\includegraphics[width=2.25in]{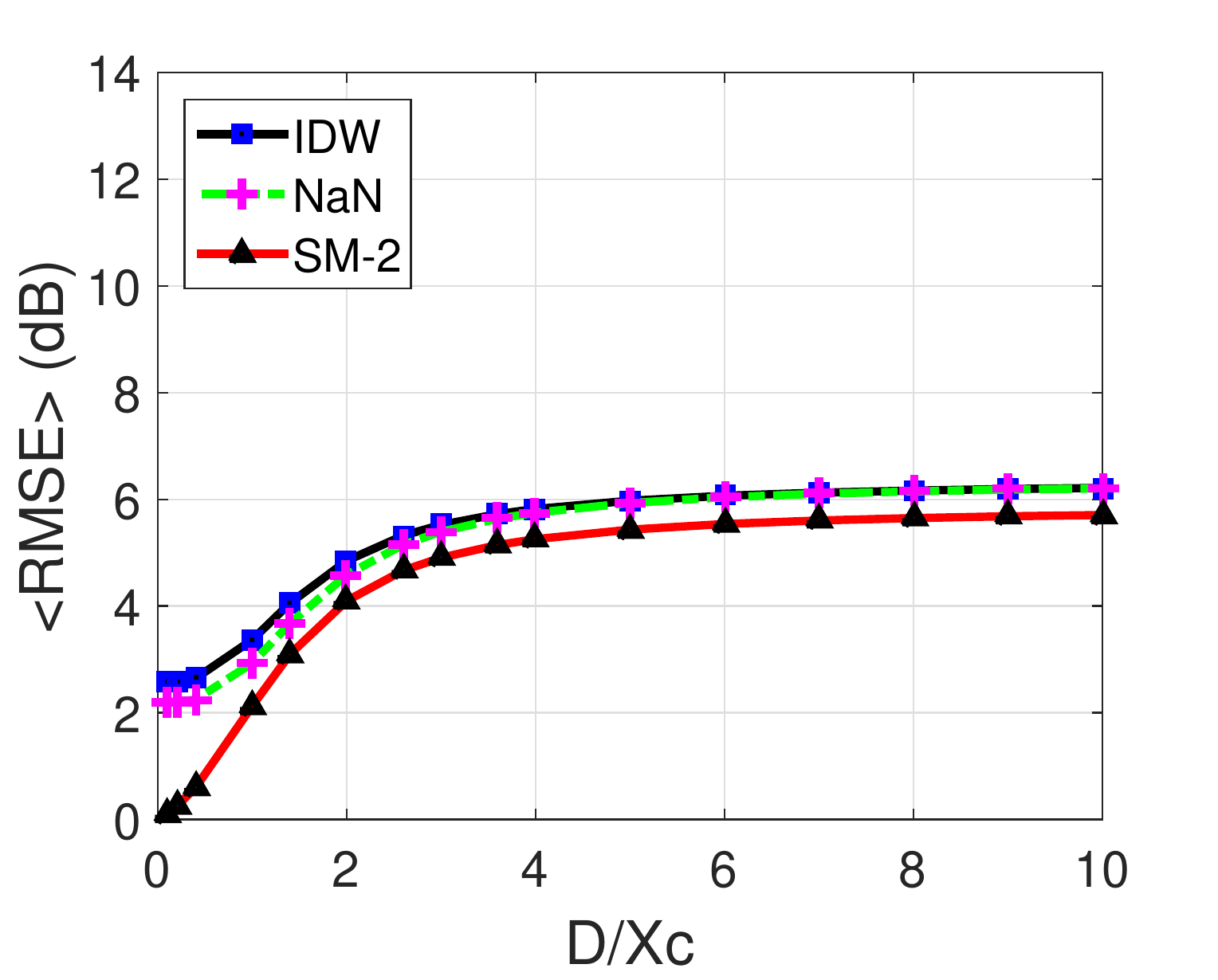}
    \label{fig:comAll_gauss_E-100_0}
	\vspace{-0.6cm}
}
\subfigure[RMSE for SM-0, SM-1, SM-2 for elliptical correlation function]{
	\includegraphics[width=2.25in]{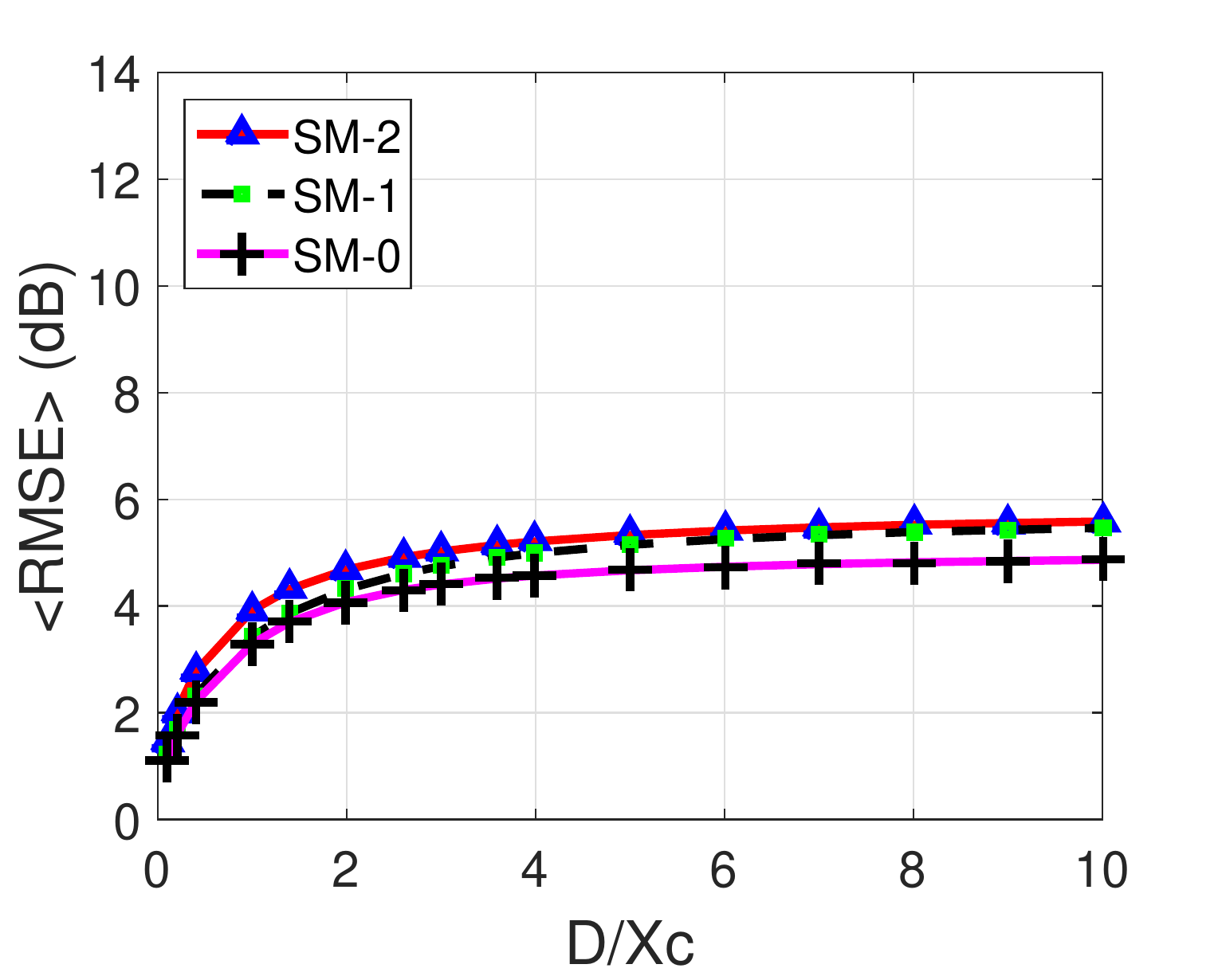}
    \label{fig:SM012_elliptical_E-100_0}
	\vspace{-0.6cm}
}
\subfigure[RMSE comparison of SM-2 with NaN, IDW for elliptical correlation function]{
	\includegraphics[width=2.25in]{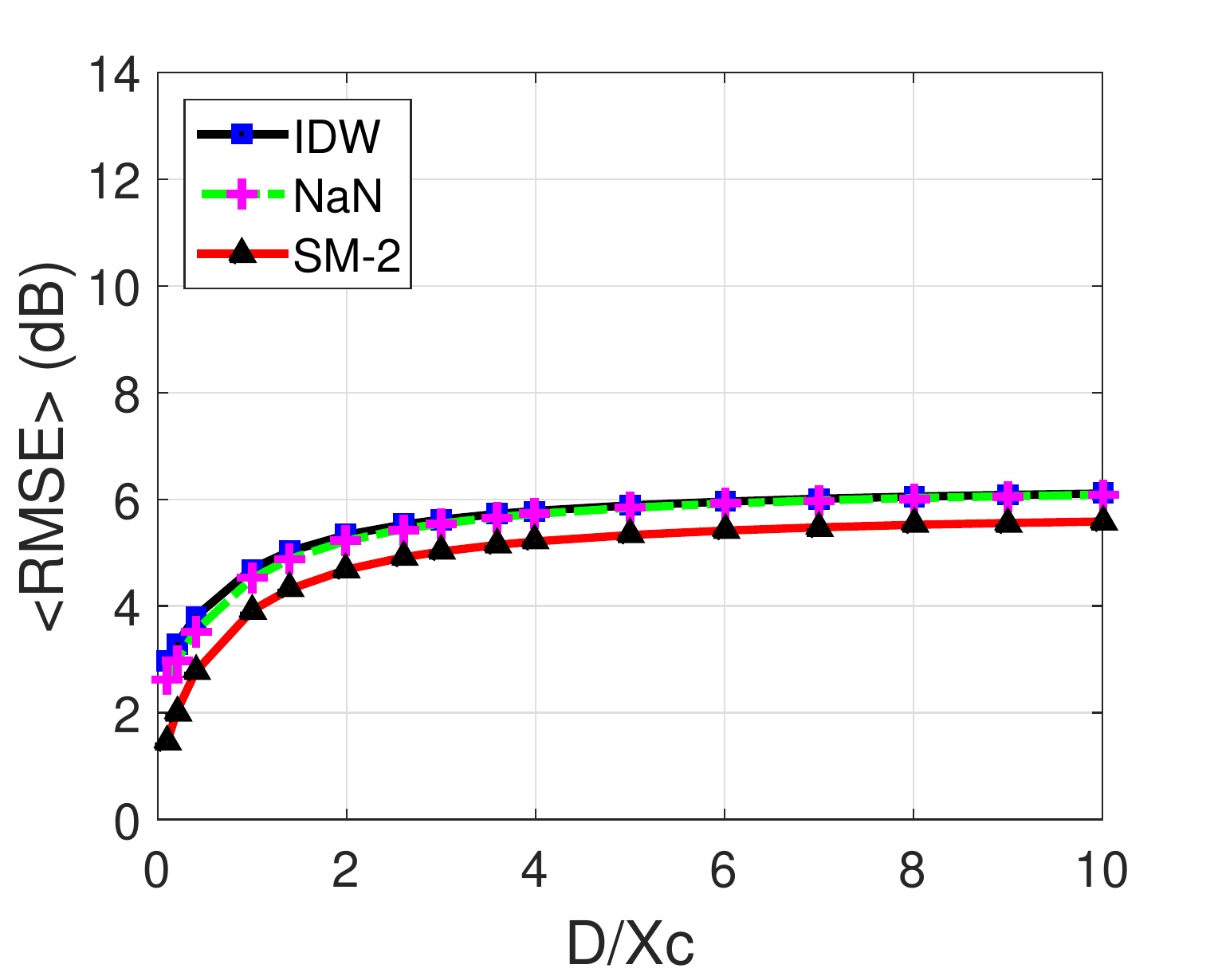}
    \label{fig:comAll_elliptical_E-100_0}
	\vspace{-0.6cm}
}
\caption[Optional caption for list of figures]{Effect of spatial correlation function on $<$\textit{RMSE}$>$ vs. $D/X_c$. $\sigma = 5$ dB and emitter at $E(-100, 0)$. %%\subref{fig:NvswoACAwACA}, \subref{fig:NvsACAgainRanFix}
}
\label{fig:RMSE_corrFn}
\vspace{-0.5cm}
\end{figure*}

In this case, the locus of constant correlation is an ellipse, tilted at some rotation angle and having unequal major and minor axes. For one possible case, with a major-to-minor axis ratio  of 3.3, we repeated the computations and obtained the results in Figs. \ref{fig:SM012_elliptical_E-100_0} and \ref{fig:comAll_elliptical_E-100_0}. Again, we see no dramatic departure from results for other correlation functions.

\eat{
\subsubsection{Elliptical correlation function}
wherein the correlation depends on direction as well as separation distance, unlike previous cases where correlation depends only on the separation distance. Taking exponential correlation expression as a base case, we represent spatial correlation between locations $a$ and $b$ as
\begin{equation}
\label{eq:auto-co-Ellip}
\begin{aligned}
c_{ab} = \sigma^2 \exp\left(-\frac{f(a,b)}{X_c}\right),
\end{aligned}
\end{equation}
For the illustration purposes, we consider the skewed ellipse where every point on ellipse has same correlation with the center of ellipse. In this case, $f(a,b) = f(a,b,L_1,L_2,\theta)$ where $L_1$ is the major axis of ellipse; $L_2$ is the minor-axis of ellipse and $\theta$ is the angle of rotation.
}

\eat{
Shadow fading $S$ is the variation about the median path loss, $P_m$, at a receiver which occurs due to large environmental obstructions. For previous results, its variability over the terrain is described by a spatial correlation function given by (\ref{eq:auto-co_1}) and (\ref{eq:auto-co}).
This expression implies that the spatial correlation between any two location with same separation distance is always same. This is not valid in the real scenarios involving obstructions such as building, trees, terrain, etc. Thus, we study performance of estimation algorithms for other correlation functions$-$
\subsubsection{Gaussian correlation function}
with spatial correlation expression as\cite{abrahamsen1997review}
\begin{equation}
\label{eq:auto-co-Gauss}
c_{ab} = \sigma^2 \exp\left[-\left(\frac{d_{ab}}{X_c}\right)^2\right]
\end{equation}
where correlation is inversely proportional to distance-squared as opposed to linear relationship in (\ref{eq:auto-co})
and
\subsubsection{Elliptical correlation function}
with spatial correlation expression as
\begin{equation}
\label{eq:auto-co-Gauss}
\begin{aligned}
c_{ab} = \sigma^2 \exp\left(-\frac{f(a,b)}{X_c}\right),\\
f(a,b) = f(a,b,L_1,L_2,\theta),
\end{aligned}
\end{equation}
where $f(a,b)$ is a skewed distance between locations $a$ and $b$ according to an elliptical shape, where $L_1$ is the major axis of ellipse; $L_2$ is the minor-axis of ellipse and $\theta$ is the angle of rotation.

Fig.~\ref{fig:RMSE_corrFn} shows the $<$\textit{RMSE}$>$ for these two cases. For the elliptical correlation function, $<$\textit{RMSE}$>$ for SM-2 is higher than SM-1 for low values of $D/X_c$, but the difference is less than 0.5 dB. We note that SM-2 consistently outperforms NaN and IDW. Error curves for NN are omitted due to the high $<$\textit{RMSE}$>$ compared to all other approaches.
}

\eat{

} 
\section{Discussion}
\label{sec:disc}
This section summarizes the major findings of our study through the analytical and graphical results for a square layout of sensors. We note that these findings can be generalized and scaled to other geometric configurations. 
\begin{itemize}
\item The ideal stochastic method (SM-0) is identical to Simple Kriging and gives the lower bound on $<$\textit{RMSE}$>$ as a function of $D/X_c$. The curve of $<$\textit{RMSE}$>$ vs. $D/X_c$ rises smoothly from (0, 0) and approaches an asymptotic value equal to $\sigma$. This error curve is independent of emitter location.

\item A key attribute of SM-0, SM-1 and SM-2 is that they estimate the parameters of the median received power, a consequence of which is that the RMS errors go to 0 as $D/X_c$ goes to 0.

\item The RMS error curves for SM-1 and SM-2 are indistinguishable from each other in all cases; thus, using the nonparametric weighting method of SM-2 incurs virtually no penalty. For large $D/X_c$, the errors are slightly higher than for SM-0; the gap depends on the geometry of the emitter location, but is never more than about 0.2$\sigma$ (1 dB for $\sigma = 5$ dB).%In the region below $D/X_c = 2$, the gap is quite small and, for some emitter locations, imperceptible.

\item IDW is computationally simpler than NaN, but NaN yields smaller RMS errors. However, the error curves for NaN and IDW do not approach 0 as $D/X_c$ approaches 0, in contrast to the SM cases. For both methods, the error curves are always above those for SM-2. %The RMS error in this limit is due to the failure of the weighted sum to estimate the true median path loss. For the results computed here, and depending on the emitter location, the error can be as large as 6 dB.

\item The NN method is an outlier in this field of comparison. While simpler than all others, it is highly sensitive to emitter location and in some cases produces very large RMS errors compared to all others.

\item SM-2 is much simpler to implement than NaN and  Ideal Kriging (SM-0), and unlike the latter, requires no knowledge of the spatial correlation function. It also provides RMS errors that are fairly uniform across the coverage error.

\item The behavior of the SM family of interpolation schemes is consistently better than IDW, NaN and NN for different spatial correlation functions.
\end{itemize} 

Taken together, it appears that SM-2  provides the best tradeoff in terms of computational simplicity, RMSE performance, and robustness to emitter location, correlation function, and other network/environment conditions. 
\section{Conclusion}
\label{sec:conc}
The construction of a radio map based on interpolation from a sparsely deployed set of distributed sensors is a promising technique for monitoring spectrum usage. The utility of such a radio map can be extended to applications such as spectrum policing, network planning and management. The currently available two weighted-sum methods, IDW and NaN, have an important `robustness' attribute in addition to not requiring knowledge of the spatial correlation function: They also do not require knowledge of the emitter location, or even knowledge of how many emitters are active; for each method, the operation is independent of this information. In the case of a single emitter of known location, the proposed stochastic methods (SM) can estimate path loss parameters to gain an advantage, and the result has been shown to be lower RMS errors in interpolating radio power. Notably, the practical approach SM-2 has low error in comparison with NaN and IDW when $D/X_c$ ranges between 0 and 1 and has consistent RMS error irrespective of the emitter location.
For the most part, however, either IDW or NaN can be used as backup interpolation methods whenever the single-emitter location, or number of emitters, is unknown. The opportunities suggested by this observation are worthy of further study.

\eat{
The two weighted-sum methods, IDW and NaN, have an important `robustness' attribute in addition to not requiring knowledge of the spatial correlation function: They also do not require knowledge of the emitter location, or even knowledge of how many emitters are active; for each method, the operation is independent of this information. In the case of a single emitter of known location, the stochastic methods (SM) can estimate path loss parameters to gain an algorithmic advantage, and the results verify this by showing lower RMS errors in interpolating radio power. However, we have also shown that the RMS errors for IDW and NaN are almost always close to those for SM-2. A notable exception is shown in Fig. \ref{fig:comAll_circular_E-100_320}, for a particular degenerate case of emitter location. For the most part, however, either IDW or NaN can be used as backup interpolation methods whenever the single-emitter location, or number of emitters, is unknown. The opportunities suggested by this observation are worthy of further study.
}

{\noindent \textbf{Acknowledgements: }  This material is based upon work supported by the National Science Foundation under ECCS-1247864 and CIF-1526908.}

\eat{
The construction of a radio map based on interpolation from a sparsely deployed set of distributed sensors is a promising technique for monitoring spectrum usage. The utility of such a radio map can be extended to applications such as spectrum policing, network planning and management. Our proposed \textit{Stochastic Methods} SM-0 and SM-1 considers environment-specific shadow-fading details and also introduces a new approach to characterize  performance as a function of the physical density $D$ of sensors along with the shadow-fading correlation distance $X_c$. Results show that these methods estimate path loss with low error, especially when $D/X_c$ ranges between 0 and 1. Such low $D/X_c$ depicts scenarios with moderate deployment density of sensors. Although these methods assume the availability of shadow-fading distribution and correlation information, we have developed an practical approach SM-2 with no requirement of such information but at the expense of some loss in accuracy. Lastly, we note that we have studied the impact of location for emitters outside the square, and a logical extension of our work will be to study the case of emitters inside the square.
} 
\appendix

\subsection{Equivalence of SM-0 and Kriging}
\label{app:equiSMKrig}
Kriging is a spatially optimal linear predictor that involves weighted averaging of samples to arrive at an estimate. Weights for Kriging are chosen according to the best linear unbiased estimator (BLUE), which yields weights that depend upon the locations of sensors used in the prediction process and on the covariannce among them. Kriging is widely studied and reported in the literature \cite{Ying2015_Incentivizing,Ying2015_Revisiting,Cressie1990_origins,Feki2008_CRNcartography,Angjelicinoski2011_Comparative} for different interpolation applications, and there are three main variations in the literature: (1) Simple Kriging, (2) Ordinary Kriging, and (3) Universal Kriging.

In our application, for the ideal case with known $A$ and $\gamma$, we will consider simple Kriging, for which the mean of the random process is assumed to be known. The optimal received power estimation at arbitrary point $0$, $Z^*(P_{r,0})$, is given by
\begin{equation}
\label{eq:simKrigPL}
Z^*(P_{r,0}) = \bm{c_0^T C_n^{-1}P_r} + (1 - \bm{c_0^T C_n^{-1}1})\mu,
\end{equation}
which parallels eq. (7) in Cressie’s tutorial paper, \cite{Cressie1990_origins}, with the following notational changes: The arbitrary point $s_0$ (point 0 in this paper)  in eq. (7) of \cite{Cressie1990_origins} is replaced by $P_{r,0}$, the received power being estimated at that point; $\bm{c}’$ is replaced by $\bm{c_0^T}$; $\bm{C}$ is replaced by $\bm{C_n}$; and $\bm{Z}$ is replaced by $\bm{P_r}$, which is an $n \times 1$ vector of the powers measured at sensors $1, 2,..,n$. In addition, Cressie’s formulation assumes a bias term, $\mu$, which is uniform throughout space, while our `bias’ term is the median path loss (of the form $A + 10\gamma \log(d_i)$), which differs at every point $i$. To accommodate this difference, we can rewrite (\ref{eq:simKrigPL}) as follows:
\eat{
where $P_r$ and $\bm{1}$ are $n \times 1$ vectors consists of collection of received power measurements at sensors and of \textit{1}s, respectively; $\mu$ is a known mean of the path loss random process which is equivalent to the median path loss defined by (\ref{eq:PLm}) as $P_{m,i} = A + 10 \gamma \log d_i$ at a given sensor $i$; and other terms are consistent with Section \ref{subsec:est_weights}. It is clear that the  at each sensor is based on $d_i$ and cannot be assumed to be a single $\mu$ value, as given by Eq.(\ref{eq:simKrigPL}). Thus, we may use the following modifications for Simple Kriging for path loss prediction as follows:
}
\begin{equation}
\label{eq:simKrigPL_rev}
Z^*(P_{r,0}) = \bm{c_0^T C_n^{-1}P_r} + (P_{m,0} - \bm{c_0^T C_n^{-1}P_m}),
\end{equation}
where $P_{m,0}$ is the median received power at point $0$; and $\bm{P_m}$ is an $n \times 1$ vector of the median powers at sensors $1, 2,..n$.

Since $\bm{P_r} = \bm{P_m}  + \bm{S}$, straightforward expansion of (\ref{eq:simKrigPL_rev}) yields
\begin{equation}
Z^*(P_{r,0}) = P_{m,0}  + \bm{c_0^T C_n^{-1} S}.
\end{equation}
From eqs. (\ref{eq:Pr0}) and (\ref{eq:mu5}) in the current paper, we see that this estimate differs from the true value of $P_{r,0}$ by just $\sigma_0 u$. Thus, the mean-square error is $\sigma_0^2$, (\ref{eq:mu5}), just as in our analysis of SM-0. We conclude that Simple Kriging and SM-0 yield equivalent (and minimum) results for mean-square estimation error.
\eat{
If we expand (\ref{eq:simKrigPL_rev}) using (\ref{eq:PL}) and (\ref{eq:PLm}), we see that (\ref{eq:simKrigPL}) and (\ref{eq:simKrigPL_rev}) are exactly same. Thus, we conclude that SM-0 and simple Kriging are equivalent in the ideal case. Further, this serves as the lower bound on path loss estimation error, and thereby allows us to use SM-0 as a baseline for non-ideal (practical) radio mapping algorithms.
}

\subsection{Estimation of $A$ and $\gamma$}
\label{app:deltaAgamma}
To estimate values $A^"$ and $\gamma$ in (\ref{eq:lo_PL}), Least-Square Estimation (LSE) assumes functional approximations as
\begin{equation}
\label{eq:funAprx}
\begin{aligned}
\bm{P}^{'} = A^{'} + 10\gamma^{'}\log_{10}\bm{d}
\end{aligned}
\end{equation}
where $\bm{P}^{'}$ is estimated received power vector of $P_1^{'},..,P_n^{'}$; and estimated values $A^{'}$ and $\gamma^{'}$ minimize the squared distance of $\|\bm{P} - \bm{P}^{'}\|$ where $\bm{P}$ is a vector of $P_1,..,P_n$.

Using LSE, solving for $A^{'}$ and $\gamma^{'}$ and using expression $P_i = A" + 10 \gamma \log d_i + S_i^"$, we get
\begin{equation}
\label{eq:delA_gamma}
\begin{aligned}
\Delta \gamma = & \gamma - \gamma^{'}
= \frac{\sum_{i=1}^n S_i \log \frac{\prod_{j=1}^n d_j}{d_i^n}}
		{10 n \sum_{i=1}^n (\log d_i)^2 - 10 (\log \prod_{i=1}^n d_i)^2},\\	
\Delta A = & A - A^{'}\\
= & \frac{\sum_{i=1}^n S_i \left(\log d_i (\prod_{j=1}^n d_j) - \sum_{j=1}^n (\log d_j)^2 \right)}
		{n \sum_{i=1}^n (\log d_i)^2 - (\log \prod_{i=1}^n d_i)^2},
\end{aligned}
\end{equation}
where both $\Delta A$ and $\Delta \gamma$ are a linear sum of shadow fading at the sensor with constant coefficient terms that are  functions of distance, $d_i$, between emitter and sensor $i$.

\eat{
\subsection{Weights for SM-1}
\label{app:SM1_weights}
In the first practical approach, SM-1, the weights are based on (\ref{eq:W})
\[\bm{W} = \bm{(c_0^T C_n^{-1})^T}\]
which are functions of correlation matrix of $\bm{S}/S_0$. But the path loss estimation algorithm (steps 1-3) uses measurements of $\bm{S^"}$ and calculates $S^"_0$. Thus, ideally, $W$ should be calculated using correlation matrices of $\bm{S^"}/S^"_0$.
For the defined local path loss model (see (\ref{eq:lo_PL})), $\bm{W}$ is not derivable. To elaborate this point, we consider a square geometry with $n = 4$ sensors as shown in Fig.~\ref{fig:sq_geo}. For each sensor, we have
\[S^"_i = S_i - \frac{S_1 + S_2 + S_3 + S_4}{4}; i = 1,..,4.\]
\begin{figure}[t]
\begin{center}
\includegraphics[width=1.15in]{pics/sq_geo.eps}
%\vspace{-1em}
\caption{An example of sensors geometry (square) for which covaraiance matrix for $\bm{S"}$ does not exits.}
\label{fig:sq_geo}
\end{center}
\vspace{-1.5em}
\end{figure}
For the correlation function given by (\ref{eq:auto-co}), each element of correlation matrix $\bm{C_4^"}$ of $\bm{s^"}$ is given as
\begin{equation}
\label{eq:corr_s"}
\bm{C_4^"}(i,j) = \left\{ \begin{array}{ll}
\frac{1}{4}(3a - 2ab - ac) & \mbox{if $i = j$};\\
\frac{1}{4}(-a + 2ab - ac) & \mbox{if $\|i - j\| = D$};\\
\frac{1}{4}(-a - 2ab + 3ac) & \mbox{if $\|i - j\| = \sqrt{2}D$},
\end{array}
\right.
\end{equation}
where
\[\begin{aligned}
a = \sigma^2, \;\; b = \exp\left(-\frac{D}{X_c}\right),\;\; c = \exp\left(-\frac{\sqrt{2}D}{X_c}\right)
\end{aligned}\]
and $i, j \in \{1,..,4\}$. The determinant of such a $\bm{C_4^"}$ is zero, and hence calculating $\bm{W}$ using $\bm{C_4^"}$ cannot be performed. This mathematical explanation can be extended to other sensor geometries and for more general number of sensors.  Therefore, we follow $\bm{W}$ derived by (\ref{eq:W}) for vector $\bm{S^"}$ as well.

From (\ref{eq:mu5}) and (\ref{eq:lo_PL}), the mean value of $s_0^{"}$ is
\begin{equation}
\label{eq:s_55}
\mu_0^{"} = \bm{W^TS} - Z_n.
\end{equation}
which is recasted as
\begin{equation}
\label{eq:s_55_est}
\mu_0^{"} = \sum_{i=1}^n w_i\bm{S}_i^" + \left(\sum_{i=1}^n w_i - 1\right)Z_n.
\end{equation}
Now the first term is a weighted sum over measured $\bm{S^"}$ values, but the second term still involves the unmeasured term $Z_n$. By restricting the optimization to be over weights whose sum is $1$, the problem problem of dealing with the unmeasured $Z_n$ disappears. 
}

\bibliographystyle{IEEEtran}
\bibliography{ref}

%\input{later_journal}
%%---------------------------------------------------------------------------%%
\end{document}